\documentclass{emulateapj}
\usepackage{natbib}

\shorttitle{Formation of Protostars}
\shortauthors{Young \& Evans}

\begin{document}

\title {Evolutionary Signatures in the Formation of Low-Mass Protostars}
\author {Chadwick H. Young \& Neal J. Evans II}
\begin{abstract}
We present an evolutionary picture of a forming star.  We assume a
singular, isothermal sphere as the initial state of the core that undergoes
collapse as described by \citet{shu77}.  We include the evolution of a
first hydrostatic core at early times and allow a disk to grow as predicted
by \citet{adams86}.  We use a 1-dimensional radiative transfer code to
calculate the spectral energy distribution for the evolving protostar from
the beginning of collapse to the point when all envelope material has
accreted onto the star+disk system.  Then, we calculate various
observational signatures ($T_{bol}$, $L_{bol}/L_{smm}$, and infrared
colors) as a function of time.

As defined by the bolometric temperature criterion, the Class 0
stage should be very short, while the Class I stage persists for much of
the protostar's early life.  We present physical distinctions among the
classes of forming stars and calculate the observational signatures for
these classes.  Finally, we present models of infrared color-magnitude
diagrams, as observed by the Spitzer Space Telescope, that should be strong
discriminators in determining the stage of evolution for a protostar.

\end{abstract}              

\section{Introduction}

In 1977, Shu presented seminal work in the theory of low-mass, isolated
star formation \citep[hereafter Shu77]{shu77}.  He presented the idea that
stars could form from inside-out collapse.  This model is still important
today because it is simple, yet it predicts so many observables in the
process of star formation.  It prescribes the evolution of inflow, the
velocity structure of the envelope, and the particular shape of the
envelope's density distribution. \citet{motte01} found that the inside-out
collapse model fit millimeter observations of protostars in Taurus and
various Bok globules.  However, these authors also found Class 0 sources in
Perseus, a less quiescent region of star formation, that had central
densities and accretion rates that were too high to be accounted for by the
Shu77 model.  Molecular line observations have also been used to compare
the predicted densities and velocities with the actual conditions in
star-forming cores.  \citet{hogerheijde00} and \citet{zhou93} presented
evidence that the envelopes of some protostars are undergoing inside-out
collapse.  Others have presented evidence against inside-out collapse.  For
example, \citet{tafalla98} suggested that L1544, a starless core, exhibits
infall motions across the entire core.  Nonetheless, it is still not clear
whether the Shu77 inside-out collapse scenario or its variants can be ruled
out.

Therefore, we have calculated the observational signatures of a star, which
is forming through inside-out collapse, so that astronomers can test this
theory in a well-defined way.  The most relevant aspects of the Shu77
inside-out collapse model are a constant accretion rate of material from
the envelope onto the star+disk system and an envelope density that is
initially described by a singular, isothermal sphere (SIS).  

The constant accretion rate in this model has given rise to the so-called
``luminosity problem'': the luminosities that result when material accretes
onto a central object with a small radius and increasing mass exceed those
seen for most young, low-mass stars  \citep{kenyon95}.  The presence of a
disk can help by increasing the accretion radius and acting as a reservoir
where matter is stored and then episodically accreted.  However, a disk
does not completely eliminate the ``luminosity problem'' in these models.

A constant accretion rate in star-forming cores should be evident when
comparing populations in the different stages of star formation.  The
transition from Class 0 to Class I object is thought to occur when the mass
of the star and disk is equal to the envelope mass.  Therefore, we should
observe equal numbers of Class 0 and I objects if they form with constant
accretion rates.  \citet{andre94} found about a 10:1 ratio for Class I and
0 sources in Ophiuchus, suggesting the Class 0 stage is very short.
However, \citet{visser02}, in an unbiased survey of dark clouds, found a
1:1 ratio.  These authors suggested that Ophiuchus has experienced a burst
of star formation in the past, resulting in the 10:1 ratio.  Future and
ongoing surveys of nearby star-forming regions will certainly offer more
information regarding the relative populations of Class 0 and Class I
sources \citep{evans03,benjamin03}.

The initial density configuration for the Shu77 model is that of a SIS,
which has $n(r)\propto r^{-2}$.  One-dimensional modelling of the
submillimeter emission from starless cores has shown that their density
distributions are well-fitted by Bonnor-Ebert spheres, but power-law density
profiles are not conclusively ruled out \citep{wardthompson94,evans01}.
Indeed, some starless cores have a density structure that seems to be
approaching the SIS. For example, three-dimensional modelling of L1544
shows a power-law density distribution with $n(r)\propto r^{-2}$
\citep{doty05}.  Other studies of the density distribution in more evolved
star-forming cores were unable to rule out the Shu77 predictions
\citep{young03}; in fact, one-third of the sources in \citet{young03} were
well-fitted by the inside-out collapse model.  For this paper, we assume
that starless cores must evolve into an SIS.  At this point, collapse
ensues, and we begin modeling the evolution of the protostar.  This work
has been expanded by \citet{lee04} to model the chemical evolution of these
star-forming cores.

Other authors have predicted the observed signatures of a
forming star.  \citet{myers98}, hereafter M98, developed a framework on
which to calculate various signatures of star formation.  Our work has been
prompted by their efforts.  However, the methods and models employed in
this work are quite different from those of M98; as a result, our
conclusions are different. We discuss these differences in this paper.

Because these models are 1-dimensional, we do not consider the role of
outflows, flattened envelopes, or asymmetric disks; therefore, we probably
underestimate the amount of short-wavelength radiation.  \citet[hereafter
W03]{whitney03} have included some of these complications. However, W03 did
not create a consistent evolutionary model but, instead, considered
``typical'' protostellar objects of different observational classes.
Fortunately, these authors were able to explore the impact of 3-dimensional
effects, which we are unable to model. In this way, we consider this effort
to be complementary to the work of W03 and compare our results to theirs.

The advent of large surveys such as 2MASS and the Spitzer Space Telescope's
Legacy programs \citep{evans03,benjamin03} provide vast sets of data
through which theories of star formation may be vigorously tested. In
this work, we hope to provide some tangible means to study the validity of
the inside-out collapse model in the role of forming stars.

\section{The Model}
 
First, we define the framework through which we have created this
evolutionary sequence.  In this section, we discuss what has been assumed
for the dust opacity, interstellar radiation field, envelope structure and
dynamics, and the evolution of the star and disk components of the system.

\subsection{Interstellar Radiation Field \& Dust Properties}

\citet{evans01} showed that, for starless cores, the interstellar radiation
field (ISRF) significantly affects both the total observed luminosity and
the shape of the observed submillimeter intensity profile; even objects
with a luminous, internal source might attribute most of their luminosity
to the ISRF \citep{young04a}. \citet{evans01} scaled the ISRF by a
constant, but we have used the opacity of \citet{draine84} dust to
attenuate the ISRF with $A_V=0.5$ (see Figure 8 in \citet{young04b}).  This
method simulates the effects of low density material in the environs of a
star-forming core.

\begin{figure}
\plotone{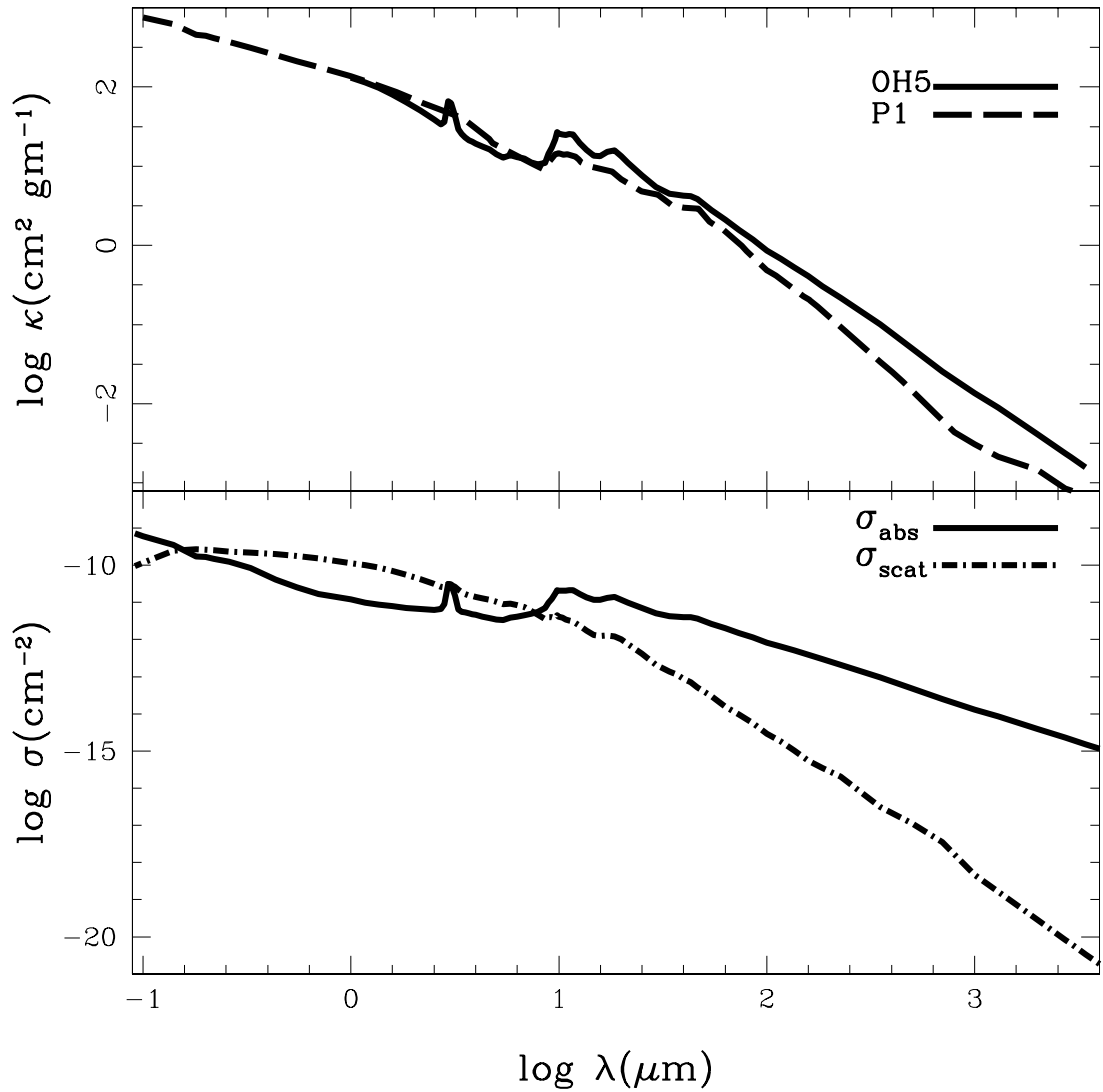} \figcaption{\label{kappa} The top plot shows
the opacity of the gas with OH5 and P1 dust (assuming a gas-to-dust ratio
of 100).  We have used the OH5 dust for our models.  However, we
used the P1 opacities for wavelengths shortward of 1 $\mu$m. The bottom
plot shows the absorption and scattering cross-sections for OH5 dust.}
\end{figure}

These authors \citep{evans01,shirley02,young03} also found that the
multi-wavelength observations were best matched by using the opacities of
the dust modeled by \citet{ossenkopf94}.  In particular, they concluded
that ``OH5'' dust, found in the fifth column of Table 1 in
\citet{ossenkopf94}, was optimal for star-forming cores.  Unfortunately,
the modeled data for OH5 dust does not include wavelengths shortward of
1.25 $\mu$m. \citet{ossenkopf94} calculated only the values for the dust
opacity ($\kappa$) and not the scattering and absorption cross-sections
($\sigma_{abs}$ and $\sigma_{scat}$) as needed by DUSTY, the radiative
transfer program we have used \citep{ivezic99,ivezic97}. Therefore, we have
obtained data from \citet{pollack94}, which includes the scattering and
absorption cross-sections for wavelengths as short as 0.091 $\mu$m.  In
Figure~\ref{kappa}, we show the opacities for OH5 dust and the opacity
calculated by \citet{pollack94} for dust grains with a radius of 0.1 $\mu$m
at a temperature of 10 K (hereafter, P1 dust); we have assumed a
gas-to-dust ratio of 100 and give the opacity of the gas in this figure.
At short wavelengths, these two types of opacities are in fairly good
agreement; unless $\tau$ is low, however, the short-wavelength opacity is
not relevant.  We used the opacity given for OH5 and $\sigma_{scat}$ for
the P1 dust to calculate the absorption coefficient for the OH5 dust.
Further, we used the albedo values given by \citet[Figure 4b]{pendleton90}
to apportion the opacity due to scattering and absorption from the 3 $\mu$m
ice feature.  Finally, we have extrapolated the cross-sections out to 3.6
cm, as required by DUSTY.  For $\sigma_{scat}$, we extrapolate by a
$\lambda^{-4}$ power-law as expected for Rayleigh scattering.  We fit the
last several data points of the OH5 absorption coefficients to determine
the $\lambda^{-1.8}$ power-law used to extrapolate $\sigma_{abs}$ out to
$\lambda=3.6$ cm.  We show the scattering and absorption coefficients in
Figure~\ref{kappa}.

\subsection{Envelope}\label{sxn-envelope}

For the density structure in the envelope, we adopt the inside-out collapse
scenario \citep{shu77}. This model begins with an SIS with a density
distribution that is proportional to $r^{-2}$.  Through some perturbation,
collapse begins inside the cloud and proceeds outward.  As collapse ensues,
the cloud's density distribution can be approximately described by a broken
power law: the inner collapsing portion, $n \propto r^{-3/2}$ (indicative
of freefall), and the outer static envelope, $n \propto r^{-2}$.  However,
there is a transition region just within the infall radius where the
density profile is significantly flatter than the $r^{-3/2}$ power law.
Therefore, we use the actual solutions to Equations 11 and 12 in
\citet{shu77}.

When the infall radius exceeds the outer radius, the Shu77 solution is
no longer valid.  Therefore, we adopt a density profile with $n
\propto r^{-3/2}$ and let the mass of the envelope (and, hence, the
fiducial density) decrease as mass is accreted onto the protostar and
disk.  

The total amount of mass is constrained by the effective sound speed
($c_s$) and the envelope's outer radius ($r_o$).  The models presented
herein all begin with cores whose initial masses are different.  We
calculate this mass from the following expression:
\begin{equation}\label{eqn-menv}
M_{env}^{t=0}=\frac{2c_s^2 r_o}{G},
\end{equation}
where $G$ is the gravitational constant, and $c_s$ is the effective
sound speed,
\begin{equation}
c_s=\left(\frac{kT}{\mu m_H}  +\frac{1}{2}v_{turb}^2\right )^{1/2},
\end{equation}  
where $k$ is Boltzmann's constant, $T$ is the isothermal temperature, $\mu$
is the mean molecular mass ($\mu=2.29$), and $m_H$ is the mass of the
hydrogen atom ($m_H=1.6733\times10^{-24}$g) and $v_{turb}$ is the turbulent
velocity (1/e Doppler width). We choose $T=10$ K and the value for
$v_{turb}$ such that the turbulent contribution to the sound speed is equal
to the thermal component ($v_{turb}=0.268$ km s$^{-1}$); $c_s=0.268$ km
s$^{-1}$.

We calculate the total envelope mass as follows,
\begin{equation}
M_{env}=\mu m_H \int\limits_{r_i}^{r_o} 4\pi r^2 n(r) dr,
\end{equation}
where $r_i$ and $r_o$ are the inner and outer radii of the envelope.  In
this paper, radii pertaining to the envelope are denoted by the lower-case
``r'' ($r_i$, $r_{inf}$, and $r_o$) while the radii of the star and disk
are denoted by the upper-case ``$R$'' ($R_\ast$, $R_i$, $R_D$).

\begin{figure}
\plotone{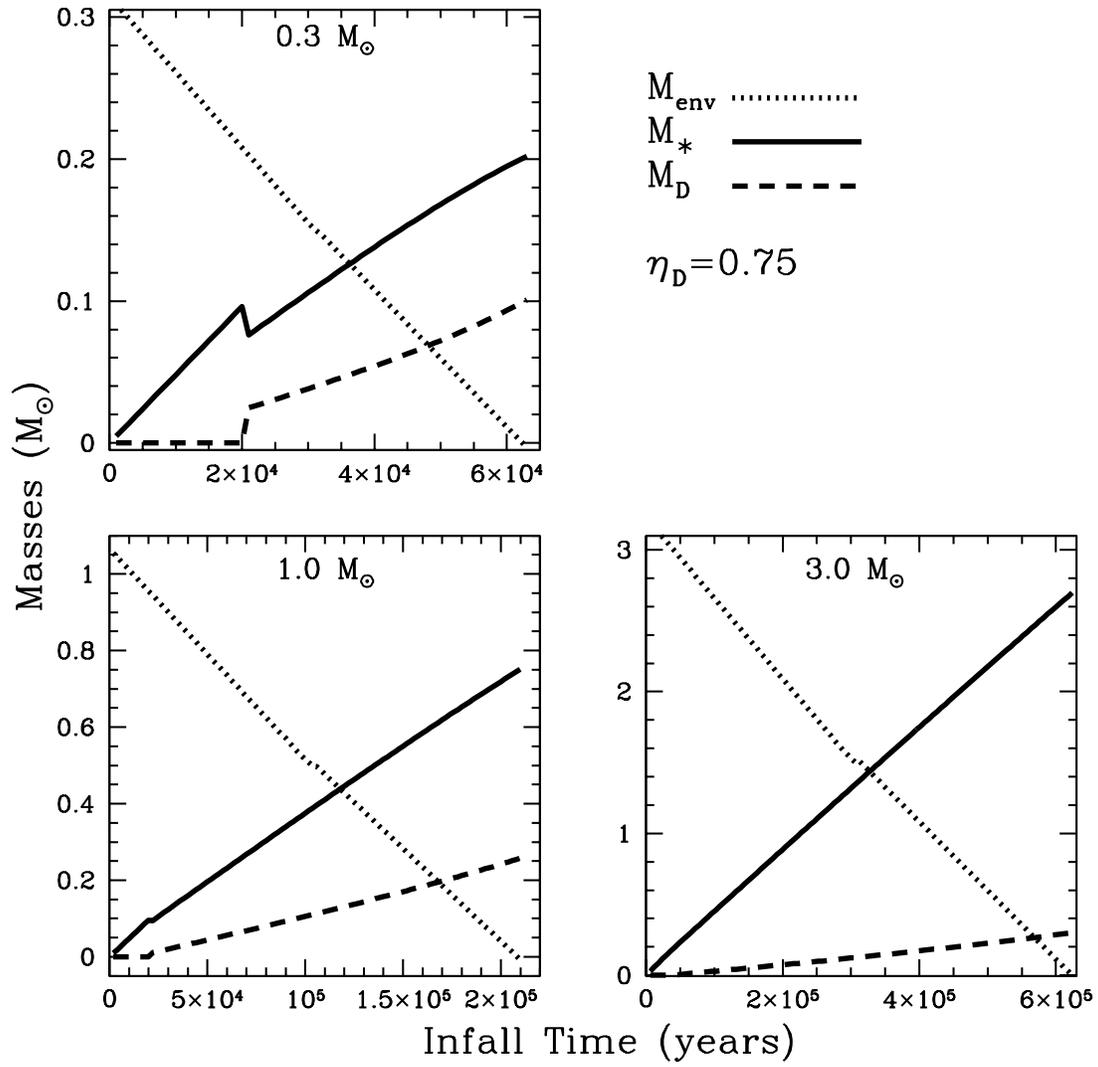} \figcaption{\label{mass} The mass of the star and disk
increase as mass from the envelope is accreted by the protostellar system
as shown here for the three mass scenarios considered. }
\end{figure}

Even though the Shu77 model has no specific mass scale, we assume that the
model is applicable to the formation of stars with different masses.  In
order to define the mass of the core, we truncate the outer radius of the
envelope, using Equation~\ref{eqn-menv}.  Such truncated envelopes are not
unheard of; \citet{motte98} found evidence for truncated outer radii for
cores in Ophiuchus. In this paper, we consider cores with three initial
masses: 0.3, 1.0, and 3.0 M$_\odot$.  With our assumed sound speed, these
masses correspond to outer radii of 1767, 5889, and 17667 AU.  We end our
modeling when all envelope material has accreted onto the star-disk system.
In these three scenarios, this event occurs at 62500, 210000, and 625000
years, respectively.  The time for collapse varies significantly among the
three models, but this variation is inherent within the inside-out collapse
model, assuming similar initial conditions.  With constant and identical
accretion rates, lower-mass objects simply form more quickly than
higher-mass objects.  We show the mass evolution for each of these cases in
Figure~\ref{mass}.

It is not clear whether this model is realistic for cores with masses as
low as 0.3 M$_\odot$. The Jean's mass for a core with $T=10$ K and density
of $10^6$ cm$^{-3}$ is $\sim$0.6 M$_\odot$ ($M_J=18$M$_\odot
T_K^{1.5}n^{-0.5}$).  However, if the cloud is cooler, it could be unstable
to collapse (e.g., if $T=5$ K and $n=10^6$, $M_J=0.2$ M$_\odot$). In fact,
much smaller mass cores can be created through turbulent fragmentation
\citep{boyd05}, so the 0.3 M$_\odot$ is probably not so absurd.
Considering that some evolved protostars are thought to be substellar
\citep{white04}, we must consider how these objects are formed.

Defining the envelope's inner radius is an issue.  One choice is that the
envelope's inner radius could be equal to the outer radius of the disk.
However, this disk radius, which is defined as the centrifugal radius
(Section~\ref{sxn-diskradius}), is very small at early times.  With a small
inner radius, the density in the inner region is unrealistically large
(e.g., $n\sim10^{10}$ cm$^{-3}$). These dense regions cause the opacity to
become very high. Further, a rotating envelope becomes flattened and
aspherical at these small radii \citep{terebey84}, so a spherical model is
not appropriate to these regions.  Therefore, we set a maximum value for
$\tau(100$ $\mu$m) and calculate the inner radius of the SIS (at $t=0$) so
that $\tau_\nu(100 \mu$m) is equal to $\tau_{max}$.
\begin{equation}\label{eqn-rienv}
r_i=\left[ \frac{\tau_{max}G2\pi}{\kappa_\nu c_s^2}+\frac{1}{r_o} \right] ^{-1} 
\end{equation}
We choose $\tau_{max}=10$ and discuss the impact of varying this in
section~\ref{sxn_tau_max}.  The envelope's inner radius follows this
formula until it is exceeded by the disk radius.  For the three cores, the
envelope's inner radius is 100 AU at the end of collapse (see
Section~\ref{sxn-diskradius}).

We define the infall rate for the case of a non-magnetic, centrally peaked
envelope density distribution as described in the \citet{shu77} scenario.
The rate of constant accretion is calculated as follows:
\begin{equation}
\dot{M}=m_o \frac{c_s^3}{G},
\end{equation}
where $m_o$ is a dimensionless constant of order unity.
Since $c_s=0.268$ km s$^{-1}$, the accretion rate, $\dot{M}$, is
$4.8\times 10^{-6}$ M$_\odot$ yr$^{-1}$.

We adopt the prescription of \citet{adams86} for the infall
rate onto the disk and star.  These authors assume that all mass is
accreted onto either the disk or star such that
$\dot{M}=\dot{M}_\ast+\dot{M}_D$ where $\dot{M}_\ast$ is the accretion rate
of envelope material directly onto the star and $\dot{M}_D$ is the accretion rate
onto the disk.  These values are calculated as follows:
\begin{eqnarray}
\dot{M}_\ast&=&\dot{M}[1-(1-u_\ast)^{1/2}],\\
\dot{M}_D&=&\dot{M}(1-u_\ast)^{1/2},
\end{eqnarray}
where $u_\ast$ is the ratio of the star and disk radii, $R_\ast/R_D$.  In
almost all cases, $u_\ast$ becomes very small in a short time and, hence,
$\dot{M}_\ast$ also approaches zero.  However, material also accretes from
the disk onto the star; this process is not included in these equations.
\citet{adams86} defined an efficiency factor---$\eta_D$, the fraction of
material in the disk that will accrete onto the star---so that the star
could gain mass and be a source of accretion luminosity.  We discuss the
implementation of this accretion process in the next section.

\subsection{Disk}

Evidence in the form of near- and mid-infrared \citep{padgett99} and
millimeter \citep{mundy96,kitamura02} observations of disks surrounding
stellar and substellar objects has recently become more convincing.  Also,
the inclusion of a disk in various models has a significant effect on the
interpretations of those models.  Therefore, to not include a disk in this
evolutionary scheme is wholly unrealistic.

We adopt the disk model developed, in theory, by \citet{adams88}, and, in
practice, by \citet{butner94}.  The density distribution for the dust
and gas (assuming homogeneous mixing) is defined as
\begin{equation}
\Sigma(R)=\Sigma_\circ\left( \frac{R}{R_f} \right) ^{-p},
\end{equation}
where $\Sigma_\circ$ is the surface density (in gm/cm$^{2}$) at $R_f$, a
fiducial radius.  We choose $p=1.5$ in accordance with the density
structure for vertical hydrostatic equilibrium \citep{chiang97}.  The mass
of the disk, given this power law distribution is given by:
\begin{equation}
M_D=\int\limits_{R_i}^{R_D}2 \pi \Sigma R dR=
\frac{2 \pi \Sigma_\circ R_f^p}{2-p}(R_D^{2-p}-R_i^{2-p}), p<2,
\end{equation}
where $R_i$, $R_D$, and $R_f$ are the inner, outer, and fiducial
radii of the disk.  

\subsubsection{Radius of the Disk}\label{sxn-diskradius}

The inner radius of the disk is the dust destruction radius defined as:
\begin{equation}\label{eqn-diskinner}
R_i=\sqrt\frac{L_\ast}{4 \pi \sigma T_{dust}^4 },
\end{equation}
where we define the dust destruction temperature, $T_{dust}=2000$ K,
and $L_\ast$ is the luminosity of the star.

For the disk outer radius, $R_D$, we adopt the centrifugal radius that
evolves with time as follows \citep{terebey84},
\begin{equation}\label{eqn-diskouter}
R_D(t)=\frac{m_\circ^3}{16}c_st^3\Omega_\circ^2,
\end{equation}
where $t$ is the time and $\Omega_\circ$ is the angular velocity of the
cloud prior to collapse; other variables are as already defined.  We set
$\Omega_\circ$ so that, at the end of the Class I stage, the disk radius is
100 AU.  These angular velocities are $1\times10^{-14}$,
$5.5\times10^{-14}$, and $3.4\times10^{-13}$ s$^{-1}$ for the 3, 1, and 0.3
M$_\odot$ models, respectively.  \citet{goodman93} found a range for
$\Omega_\circ$ from $9.7\times10^{-15}$ to $1.3\times10^{-13}$ s$^{-1}$.
The upper end of this range is about one-third of $\Omega_\circ$ for the
0.3 M$_\odot$ model, but the least massive of the cores in
\citet{goodman93} was 0.6 M$_\odot$. We assume that less massive and
smaller cores have higher angular velocities.

\subsubsection{Mass of the Disk}\label{sxn-diskmass}
\begin{figure}
\plotone{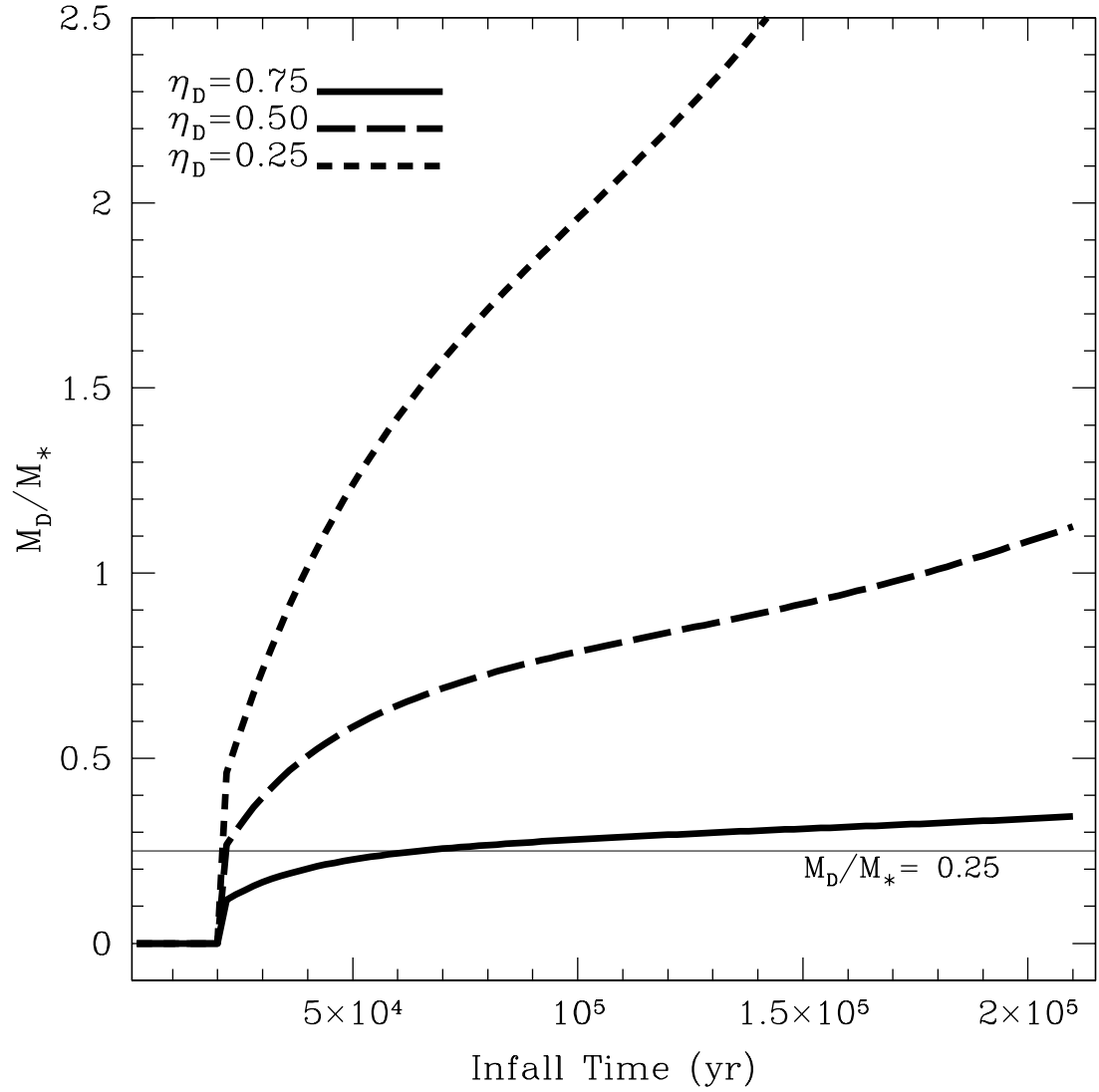} \figcaption{\label{mratio}We show the ratio
of disk to stellar mass for the 1 M$_\odot$ model as it evolves with
time. We have chosen $\eta_D=0.75$ so that this ratio approaches 0.25. For the 0.3
and 3.0 M$_\odot$ models, we use $\eta_D=0.7$ and $\eta_D=0.9$, respectively.}
\end{figure}

We evolve the mass of the disk via the expression given by \citet{adams86}:
\begin{equation}\label{eqn-mdisk}
M_D=(1-\eta_D) \int_{t_o}^t \dot{M_D}dt=(1-\eta_D)\mathcal{M_D} M,
\end{equation}
where $t_o$ is the time when $u_\ast=1$ (i.e., $R_D=R_\ast$).  We assume
$t_o=0$. Further, \citet{adams86} defines $\mathcal{M_D}$ as follows:
\begin{equation}\label{script_m} 
\mathcal{M_D}=\frac{1}{3}u_\ast  \int_{u_\ast}^1 (1-u_\ast)^{1/2}u^{-4/3}du.
\end{equation}
We evaluate this expression numerically as suggested by \citet{adams86}.
Finally, $M$ is the total mass accreted (i.e. $M=\dot{M}t$).
\citet{adams86} give $\eta_D$ as a free parameter; $\eta_D$ is the fraction
of material accreted onto the disk that will eventually accrete onto the
star.  To determine a value for $\eta_D$, we assume that the ratio of star
to disk mass must be $\sim1/4$, which is in accord with theoretical work
\citep{li02}.  In Figure~\ref{mratio}, we show this ratio for several
values of $\eta_D$.  We choose $\eta_D=0.75$ so that $M_D/M_\ast\sim1/4$
for the 1 M$_\odot$ model.  We apply the same criterion for the 0.3 and 3
M$_\odot$ models and set $\eta_D=0.7$ and $\eta_D=0.9$, respectively.  In
section~\ref{sxn-parameters}, we explore the effects of allowing it to
vary.

\subsubsection{Luminosity of the Disk}\label{sxn-diskluminosity}

The temperature distribution for the disk is defined by the following,
\begin{equation}\label{eqn-tdisk}
T(R)=T_\circ\left( \frac{R}{R_f} \right) ^{-q} (K),
\end{equation}
where $T_\circ$ is the temperature at the fiducial radius, $R_f$.  We set
$q=0.5$, a temperature distribution that decreases more slowly with radius
than expected for a flat disk, to simulate flaring and disk accretion
\citep{butner94,kenyon87}.

The luminosity of the disk has several components as given by
\citet{adams86}.  First, envelope material falling onto the disk will act
as a source of luminosity.  Then, there is a source of ``mixing
luminosity'' that arises, basically, from the mixing of newly accreted
material and that material already in orbit around the star.  Finally,
\citet{adams86} assumed that some fraction ($\eta_D$) of the disk material
will frictionally dissipate its remaining orbital energy and fall onto the
star.  \citet{adams86} give the expression for $L_{acc}^D$ in equation 33b
of their work; we use this equation for $L_{acc}^D$, which includes all
three of these components, in the evolutionary model. As the disk radius
grows larger with time and $u_\ast \rightarrow 0$, their expression for
$L_{acc}^D$ simplifies to \citep{adams87}
\begin{equation}\label{eqn-ldisk-approx}
L_{acc}^D\approx\frac{1}{2} \eta_D L_\circ,
\end{equation}
where $L_\circ \equiv \frac{G M_\ast \dot{M}_\ast}{R_\ast}.$

\subsection{Star}

For the star, we define several parameters: mass ($M_\ast$), luminosity
($L_\ast$), radius ($R_\ast$), and effective temperature ($T_{eff}$).  Each
of these quantities evolve over time, and they are generally
interdependent.

\subsubsection{Mass of the Star}

For the stellar mass, we subtract the disk mass from the total accreted
mass as in \citet{adams86}:
\begin{equation}\label{eqn-mstar} 
M_\ast=M-M_D=[1-(1-\eta_D)\mathcal{M_D}]M.
\end{equation}
All variables are as previously defined. Included in this equation are two
means by which the star gains mass.  Material accretes directly from the
envelope onto the star until the disk grows in size and accretes most of
the infalling envelope mass.  Then, the star mostly gains material that has
accreted onto and is processed through the disk.

\subsubsection{Luminosity of the Star}\label{sxn-starlum}

The luminosity of the star has  two components: that arising from
accretion ($L_{acc}^\ast$), the dominant source of luminosity at early
times, and the luminosity due to gravitational contraction and
deuterium burning ($L_{phot}$).  Simply, $L_\ast=L_{acc}^\ast+L_{phot}$.  

\citet{adams86} calculate the accretion luminosity from material accreting
onto the star.  Their calculations include the luminosity from material
that falls onto the star and the energy released due to differential
rotation of the protostar.  \citet{adams86} give an expression for
$L^\ast_{acc}$ in equation 33a of their work; we use this prescription for
L$_{acc}^\ast$. As the disk radius grows larger with time and $u_\ast
\rightarrow 0$, this simplifies to \citep{adams87}
\begin{equation}\label{eqn-lacc-approx}
L_{acc}^\ast\approx\frac{1}{2} \eta_D^2 \eta_\ast  L_\circ.
\end{equation}

\citet{adams86} define $\eta_\ast$ as an ``efficiency factor'' that
dictates how the star dissipates the energy due to differential rotation.
\citet{adams87} considered this value to be a free parameter but chose
$\eta_\ast=0.5$ for their standard model, which we use as well.  In
section~\ref{sxn-parameters}, we discuss the effects of varying
$\eta_\ast$.

$L_{phot}$ was calculated by \citet[hereafter DM]{dantona94}.  We have made
linear fits to their pre-main sequence tracks with opacities from
\citet{alexander89} (Tables 1 and 5 of DM).  First, we have fit a power-law
in the luminosity-time plane for each stellar mass given by DM.  For masses
less than 0.2 M$_\odot$, where a single power-law is not appropriate, we
have fit two-piece power-laws to DM's data.  For times earlier than those
covered by DM's tracks, we have assumed a power-law expression:
$L_{phot}=L_\circ^{phot} \left(\frac{t}{t_\circ}\right)^5$, where $t_\circ$
is the earliest time in the calculations by DM, and $L_\circ^{phot}$ is the
luminosity of the pre-main sequence star at time $t_\circ$.  This equation
is ad hoc and meant to smoothly bridge the transition from where there is
no data for L$_{phot}$ to the point where DM's evolutionary tracks begin.
Finally, to obtain the appropriate value for $L_{phot}$, we linearly
interpolate between masses for a given time in the star's evolution.

As noted in M98, the beginning of infall and accretion luminosity is not
the same time as that for the onset of the luminosity represented by
$L_{phot}$.  We adopt, as M98 did, a difference in these two timescales of
$10^5$ years such that, for $L_{phot}(t)$, we take $t=t_{phot}+10^5$ yr for
particular values of $t_{phot}$ as given by DM; this assumption is based on
the theoretical work of \citet{stahler83}.  After collapse begins, the
forming star must wait $10^5$ years before the luminosity due to
contraction and deuterium burning (as described by DM) will begin.
Further, the luminosity from DM's models does not become significant until
$\sim7\times10^4$ years.  Therefore, $L_{phot}$ is not relevant until $t\sim
1.7\times10^5$ years, which is greater than the time required for the 0.3
M$_\odot$ to collapse and only $4\times10^4$ years more than the collapse
time for the 1 M$_\odot$ core.

With this final term, the total luminosity of the protostellar system is
now made of three components: $L_{acc}^\ast$, the luminosity due to
accretion and differential rotation of the protostar; $L_{phot}$, the
luminosity arising from gravitational contraction and deuterium burning in
the protostar, and $L_{acc}^D$, the luminosity from accretion onto the disk
and dissipation of the orbital energy within the disk.  

\begin{figure}
\plotone{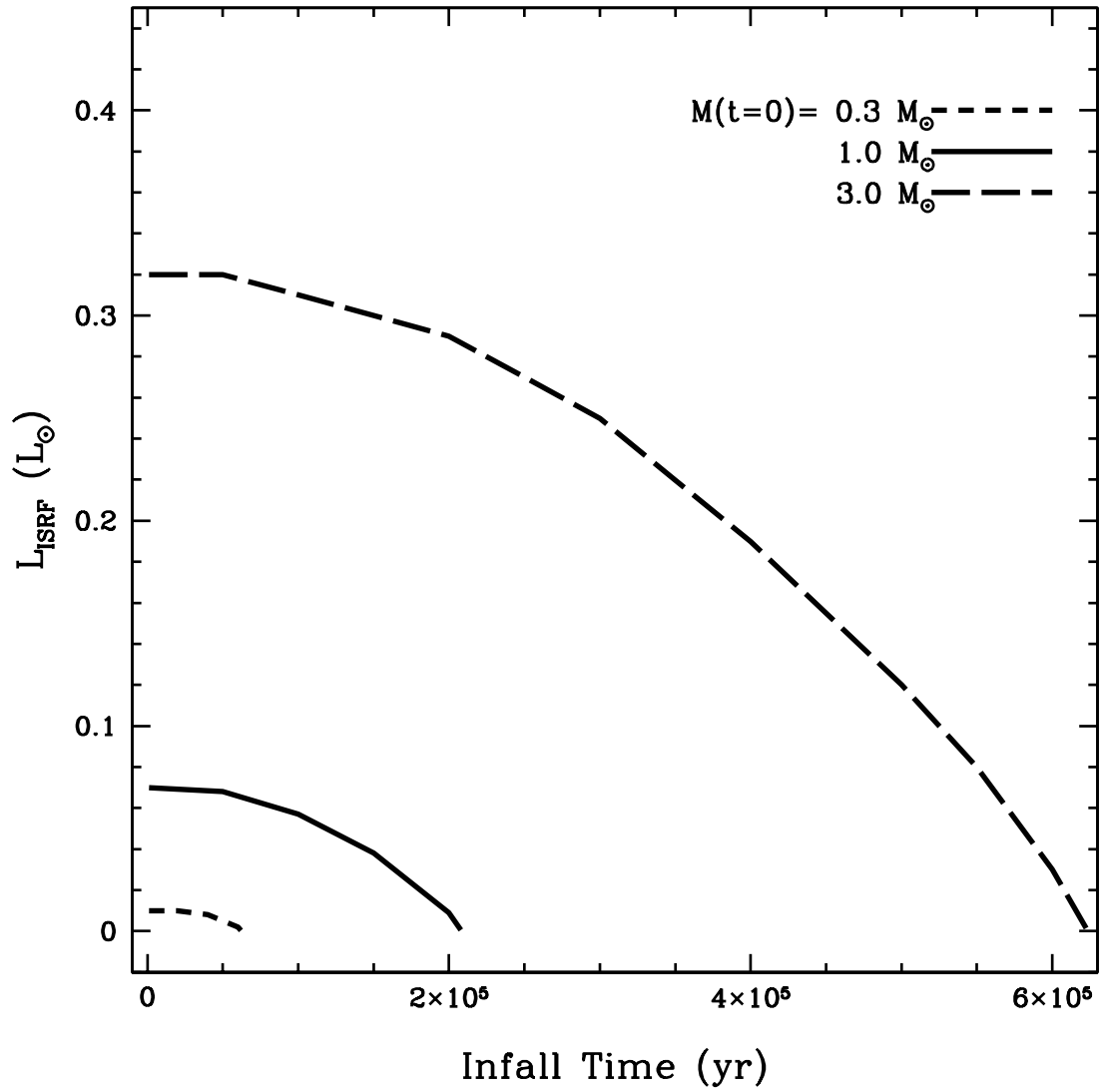} \figcaption{\label{lisrf}Evolution of
the luminosity from the ISRF with time.  Each line represents a core
with a different initial mass.}
\end{figure}

In addition, there is the luminosity that results from the ISRF,
$L_{ISRF}$.  We calculated $L_{ISRF}$ by illuminating a core that has no
internal source.  However, the core does have an evolving density
distribution identical to the three mass scenarios presented herein.  Then,
we calculate the luminosity that results from the dust grains, which are
heated externally.  At early times, the external radiation field
contributes most of the luminosity. As the envelope mass decreases and the
internal source luminosity grows, $L_{ISRF}$ becomes insignificant.  We
plot the evolution of $L_{ISRF}$ in Figure~\ref{lisrf}.

In conclusion, the total luminosity is given as
\begin{equation}
L_{tot}=L_{acc}^\ast+L_{phot}+L_{acc}^D+L_{ISRF}.
\end{equation}

For $T_{eff}$, we use the Stefan-Boltzmann Law,
\begin{equation}\label{eqn-teff}
T_{eff}=\left(\frac{L_\ast}{\sigma 4 \pi  R_\ast^2}\right)^{1/4},
\end{equation}
where $L_\ast=L_{acc}^\ast+L_{phot}$.  At early times, the effective
temperature is very low, $\sim100$ K because the radius of the first
hydrostatic core (Section~\ref{sxn-fhc}) is $\sim$5 AU.  When the stellar
radius approaches 2-5 R$_\odot$, $T_{eff}$ becomes more stellar-like ($\sim
3000$ K).

\subsubsection{Radius of the Star - Simulating the FHC}\label{sxn-fhc}

We allow the radius of the star to evolve as suggested by \citet[see Figure
1 in their paper]{palla91}.  In their calculations, they find that the
radius of the star rises to about 2-5 solar radii; this result is in accord
with the assumption of a constant radius at 3 R$_\odot$ by M98.  However,
the time at which to apply these calculations is not so clear.

\begin{figure}
\plotone{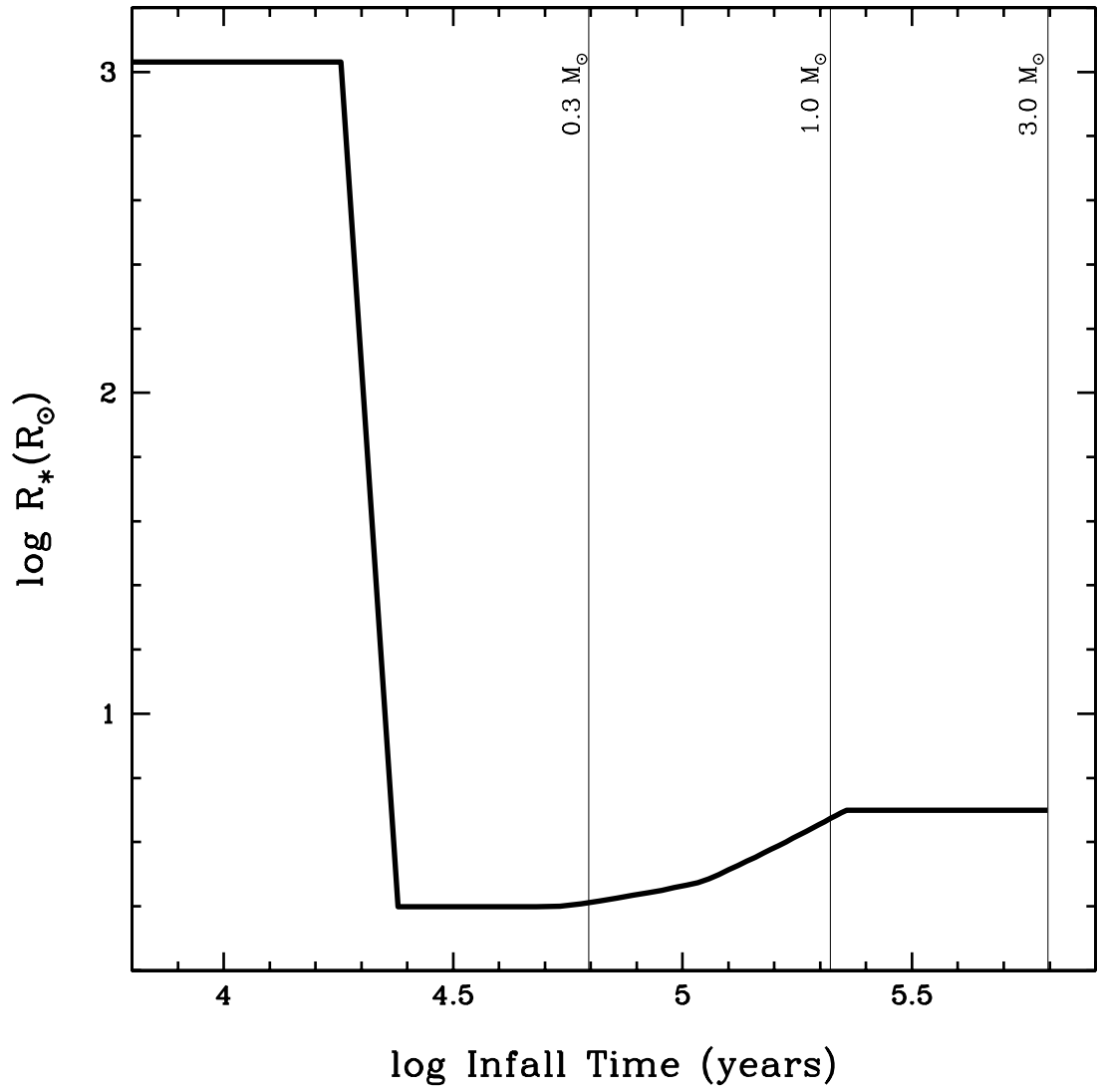} \figcaption{\label{rstar} We show the data
used for the radius of our evolving protostar.  The radius in the first
20,000 years simulates the first hydrostatic core phase.  The radius for
the remaining time is from Figure 1 of Palla \& Stahler (1991). The thin,
vertical lines show where the evolution ends for the various masses
modeled.}
\end{figure}

In the evolution of a protostar, the early years are occupied by the first
hydrostatic core.  While not yet clearly observed, the first hydrostatic
core has been predicted \citep{boss95,masunaga98}.  \citet{boss95}
concluded, based on some simple arguments, that the lifetime of this stage
should be short, only about 20,000 years.  Further, \citet{masunaga98} have
determined that the average radius of this core should be about 5 AU.  The
transition between this very large core and the smaller core described by
the calculations of \citep{palla91} is not well-understood.  We have
assumed the stellar radius evolves as shown in Figure~\ref{rstar}.  In the
beginning, the radius of the first hydrostatic core is 5 AU.  At t=20,000
years, we allow the radius to decrease from 5 AU to the radius calculated
by \citet{palla91}.  This transition lasts 100 years and is described, in
our model, by:
\begin{eqnarray}
R_\ast (AU)=5\left [1-\left (\frac{t-20000}{100}\right )^{0.5}\right
]+R_\ast^{PS}, \\
20,000<t<20,100 \nonumber
\end{eqnarray}
where $t$ is the time in years and $R_\ast^{PS}$ is the value for the
radius calculated by \citet{palla91}.  This equation is somewhat ad hoc and
simply used to simulate the transition between the large radius as predicted
by \citet{masunaga98} and the much smaller radius of the actual star as
predicted by \citet{palla91}.

\begin{figure}
\plotone{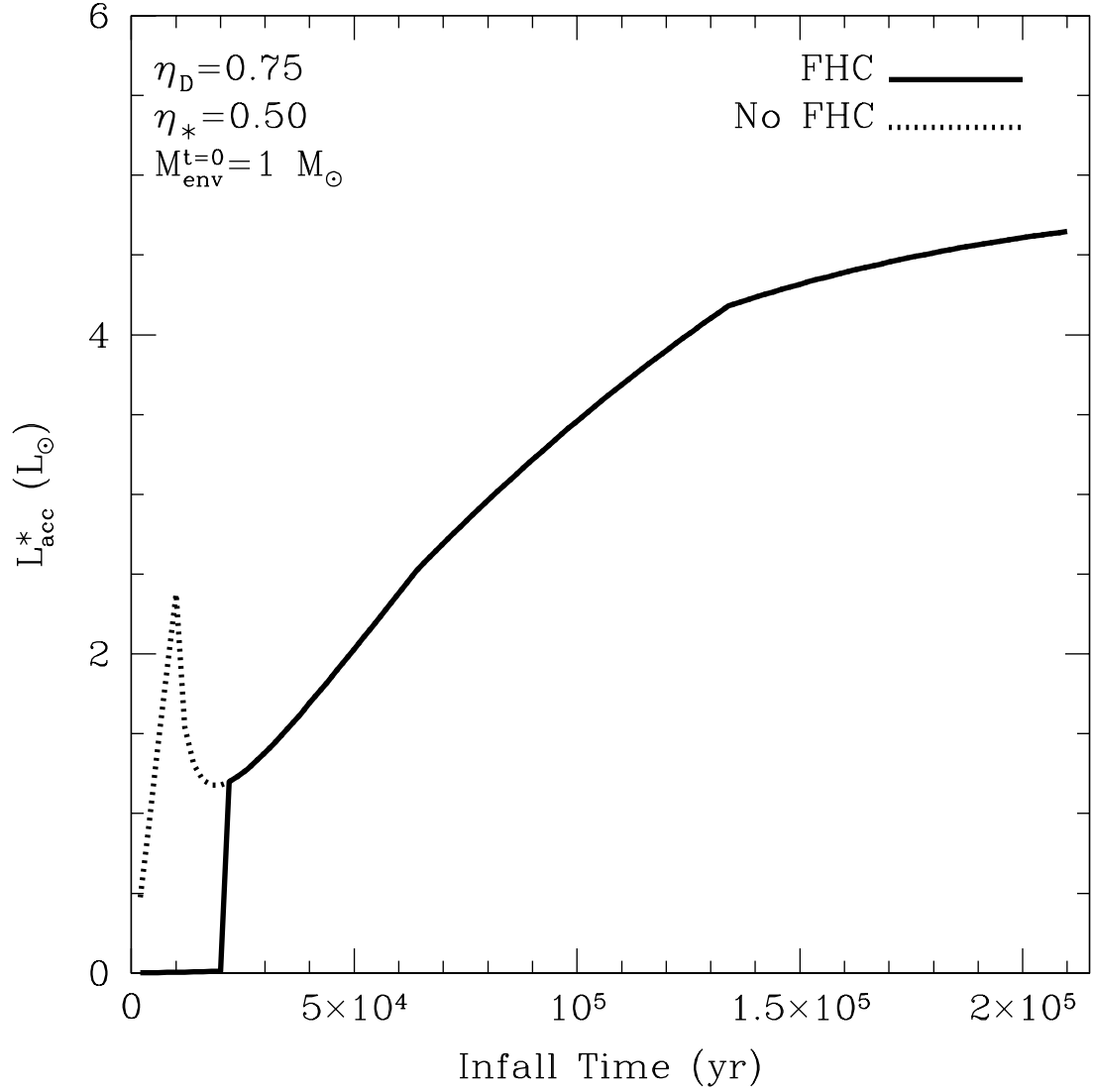} \figcaption{\label{fhc} The solid line represents
$L_{acc}$ as it evolves with time; the first hydrostatic core (FHC) is
included.  Without the FHC, $L_{acc}$ evolves as represented by the dotted
line. For our models, we include the FHC.}
\end{figure}

There are consequences for including this large radius at early times.
Because the centrifugal radius is very small and, hence, the disk has not
formed, the luminosity at these early times is derived wholly from
spherical accretion onto the central source.  If this central source is
small, the accretion luminosity can be very high.  In Figure~\ref{fhc}, we
show the evolution of the accretion luminosity for two scenarios.  In one
case, we have included the FHC.  We also plot $L_{acc}^\ast$ when there is
no FHC.  In this case, the stellar radius evolves via the data of
\citet{palla91} (i.e., between 2 and 5 R$_\odot$).  Without a FHC, the
accretion luminosity rises quickly because there is no disk and material is
accreting directly onto the star.  At about $2\times10^4$ years, the
centrifugal radius has increased so that a disk can form. Then, the
accreting material is processed by the disk causing $L_{acc}^\ast$, which
arises from accretion onto the star, to decrease.

In summary, we let the mass of the star and the accretion luminosity evolve
as defined by \citet{adams86}, the radius of the star change as predicted by
\citet{palla91} (except at early times), and the luminosity due to
deuterium burning and gravitational contraction of the PMS star evolve as
calculated by DM.  The effective temperature, given these other factors, is
defined by the Stefan-Boltzmann Law.

\section{Signatures}\label{sxn-signatures}

In this section, we discuss the various observational signatures in
the evolution of protostellar systems.

We calculate the bolometric temperature by the prescription given in
\citet{myers93},
\begin{equation}
T_{bol}\equiv[\zeta(4)/4\zeta(5)]h\bar{\nu}/k=1.25\times10^{-11}\bar{\nu}, 
\end{equation}
where $\zeta(m)$ is the Riemann zeta function of argument $m$, $h$ is
Planck's constant, $k$ is Boltzmann's constant, and the mean
frequency, $\bar{\nu}$, is the ratio of the first and zeroth frequency
moments:
\begin{equation} 
\bar{\nu}\equiv I_1/I_0 , I_m=\int\limits_{0}^{\infty} \nu^m S_\nu d\nu.
\end{equation}

In addition, we calculate the bolometric luminosity:
\begin{equation}\label{eqn-lbol}
L_{bol}=\int\limits_{0}^{\infty} 4 \pi D^2 S_\nu d\nu
\end{equation}
where $S_\nu$ is the flux density, and $D$ is the distance. We set $D=140$
pc, suitable for nearby star-forming regions. We also calculate
$L_{bol}/L_{smm}$ for our models \citep{andre93}.  $L_{smm}$ is found by
integrating Equation~\ref{eqn-lbol} from 350 $\mu$m to $\infty$.
 
We also calculate the fluxes that would be seen by the photometric bands on
the Spitzer Space Telescope by convolving the modeled spectral energy
distribution (SED) with the bandpasses for the MIPS instruments.  These
fluxes do not vary substantially from the monochromatic fluxes for the
central wavelength of each bandpass.  To convert the fluxes to magnitudes,
we use these zero-point fluxes for MIPS bands 1-3 (24, 70, and 160 $\mu$m),
respectively: 7.2, 0.8, and 0.17 Jy \citep{young05}.  We do not convolve
the MIPS resolution element with the model but assume that all emission is
included within this beam.

\section{Radiative Transfer}

We use the radiative transfer code, DUSTY, as developed by \citet{ivezic99}
to calculate the temperature distribution in the envelope and the emergent
SED of the star, disk and envelope.  In this section, we discuss the effect
and treatment of scattering by dust grains in these calculations.

DUSTY assumes that scattering from dust grains is isotropic. Longward of 10
$\mu$m, the SED is barely affected because
the scattering cross-section ($\sigma_{scat}$) is significantly less than
the absorption cross-section ($\sigma_{abs}$, see Figure~\ref{kappa}).
Further, the effect is also minimal at wavelengths shortward of 10 $\mu$m
when the interstellar radiation field is not included.

\begin{figure}
\plotone{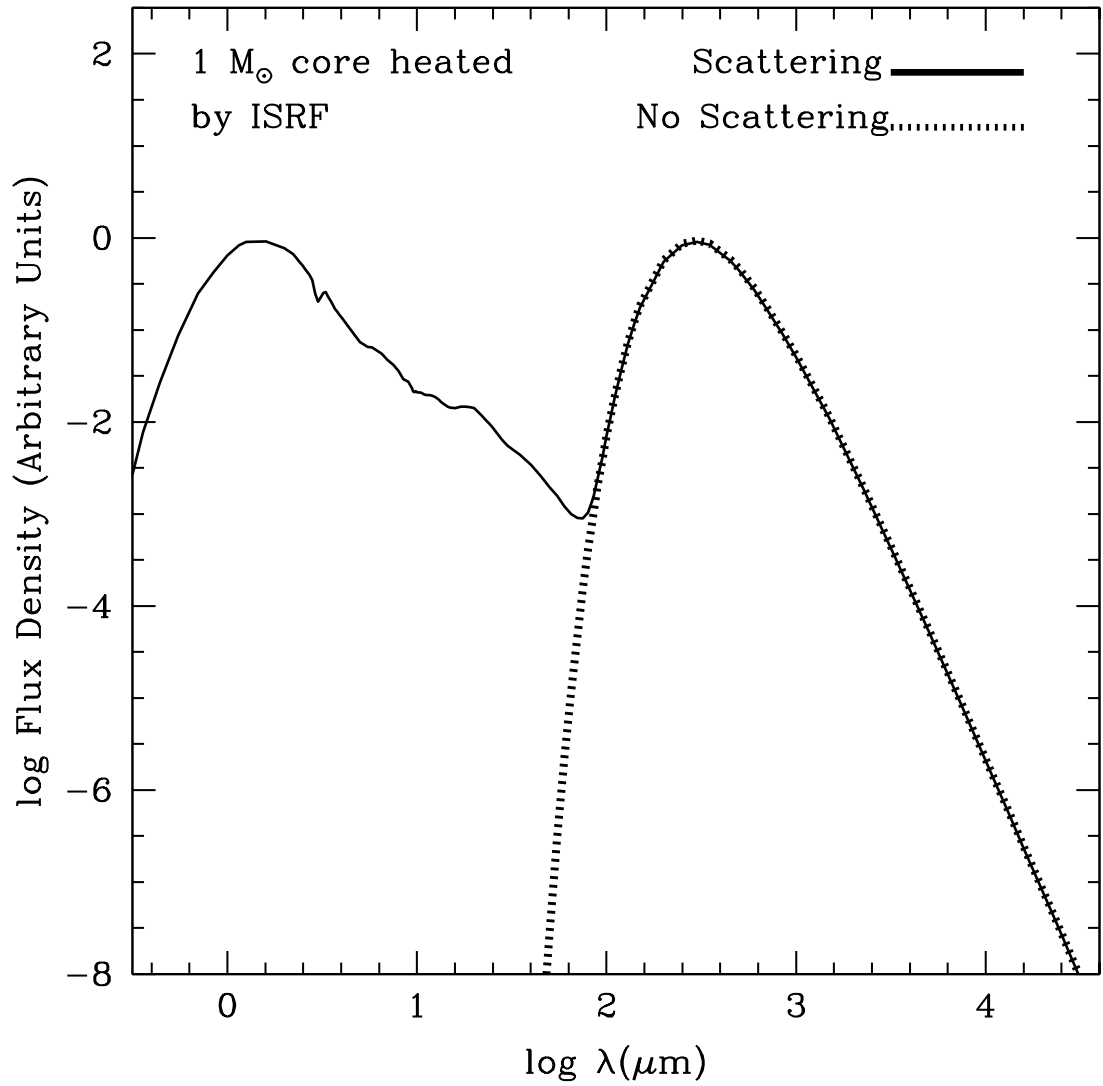} \figcaption{\label{scat_isrf} We plot
the SED of a 1 M$_\odot$ core (with $\tau_{100\mu m}=1$).  This model does
not include a central luminosity source; the heating is entirely due to the
ISRF.  The solid line includes the effects of isotropic scattering while
the dotted line does not.}
\end{figure}

Unfortunately, the assumption of isotropic scattering causes some problems
when the interstellar radiation field is included.  In
Figure~\ref{scat_isrf}, we show the SED for a core with a mass of 1
M$_\odot$ (with $\tau_{100\mu m}=1$).  In these models, the only heating of
the dust grains is externally from the ISRF.  For the solid line, we
include the effects of isotropic scattering by the dust grains while the
dashed line shows the SED without scattering. Both SEDs have a peak at
submillimeter wavelengths as is expected.  However, the SED with scattering
included also has a peak in the near-infrared.  Of course, we do not
observe strong near-infrared radiation from starless cores.  At short
wavelengths, these dust grains preferentially forward scatter light, so
neglecting the anisotropic nature of the scattering causes this unrealistic
flux in the near-infrared.

Our options are either a) neglect the effects of scattering in calculating
the emergent SED or b) ignore the ISRF.  The latter is not really feasible
because, at early times, the ISRF provides the sole source of heating and,
hence, ignoring it radically affects the temperature profile for the core.
Therefore, we opt for the first alternative and ignore the effects of
scattering.

\section{An Evolving Protostar}

\begin{figure}
\plotone{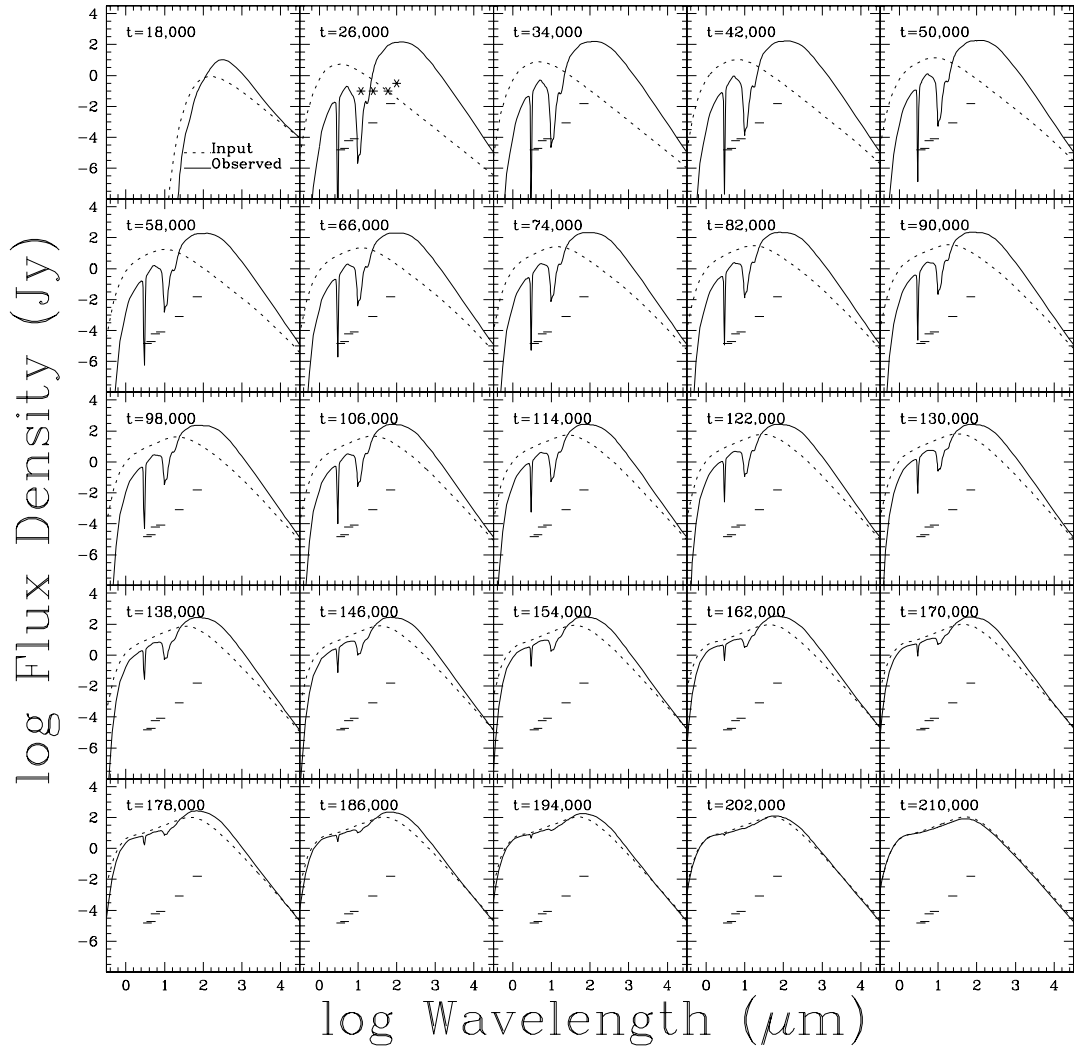} \figcaption{\label{sed} Each panel in the
plot has the SED for a core with initial mass of 1 M$_\odot$, a temperature
of 10 K, $\tau_{max}=10$, $\eta_D=0.75$, and $\eta_\ast=0.5$.  The age for
the model is labeled in each panel. The dotted line is the input disk
spectrum, and the solid line is the observed SED as calculated by DUSTY.
Sensitivities for the Spitzer Cores to Disks Legacy program are shown as
horizontal bars; IRAS sensitivities are in the second frame as
asterisks. The distance is 140 pc. }
\end{figure}

With these methods and assumptions, we have calculated the SED for an
evolving protostar. In Figure~\ref{sed}, we show the SED for particular
times in the evolution of the core that began with a pre-collapse mass of 1
$M_\odot$. We have also set $\tau_{max}=10$, $\eta_D=0.75$, and
$\eta_\ast=0.5$ (see sections~\ref{sxn-envelope},~\ref{sxn-diskmass},
and~\ref{sxn-starlum}). The solid line is the emergent SED as observed at
140 pc, the distance to Taurus. The dashed line represents the star+disk
SED; this spectrum is for the central source and is the input for DUSTY.
The bars represent the sensitivity for the Spitzer Space Telescope c2d
Legacy program \citep{evans03}; we have increased the 70 $\mu$m sensitivity
by a factor of three based on in-flight performance.  The asterisks in the
second frame are IRAS sensitivities.

We calculate the observational signatures, described in
Section~\ref{sxn-signatures}, for models with different initial conditions
and whose evolution proceeds in different ways. We use different timesteps
for the models: $\Delta t=1000$, 2000, and 6000 years for the 0.3, 1, and 3
M$_\odot$ models, respectively.  These timesteps are each about 1\% of the
total infall time.

\section{Free Parameters}\label{sxn-parameters}

In this section, we explore the effects of various parameters in these
models.  We show how the model changes when we use different values for
$\eta_D$, $\eta_\ast$, and $\tau_{max}$. We find that neither $\eta_D$ or
$\eta_\ast$ have a large effect on the observational signatures of the
SEDs, but the choice for $\tau_{max}$ can significantly affect the
short-wavelength emission during all stages of evolution. We adopt these
values: $\eta_D=0.75$, $\eta_\ast=0.5$, and $\tau_{max}=10$.

\subsection{$\eta_D$}

This factor determines what fraction of the disk material will dissipate
its energy and accrete onto the star.  It is relevant in the calculation of
the disk mass ($M_D$, equation~\ref{eqn-mdisk}), the disk accretion
luminosity ($L_{acc}^D$, equation~\ref{eqn-ldisk-approx}), the mass of the star
($M_\ast$, equation~\ref{eqn-mstar}), and  the stellar accretion
luminosity ($L_{acc}^\ast$, equation~\ref{eqn-lacc-approx}).  

\begin{figure}
\plotone{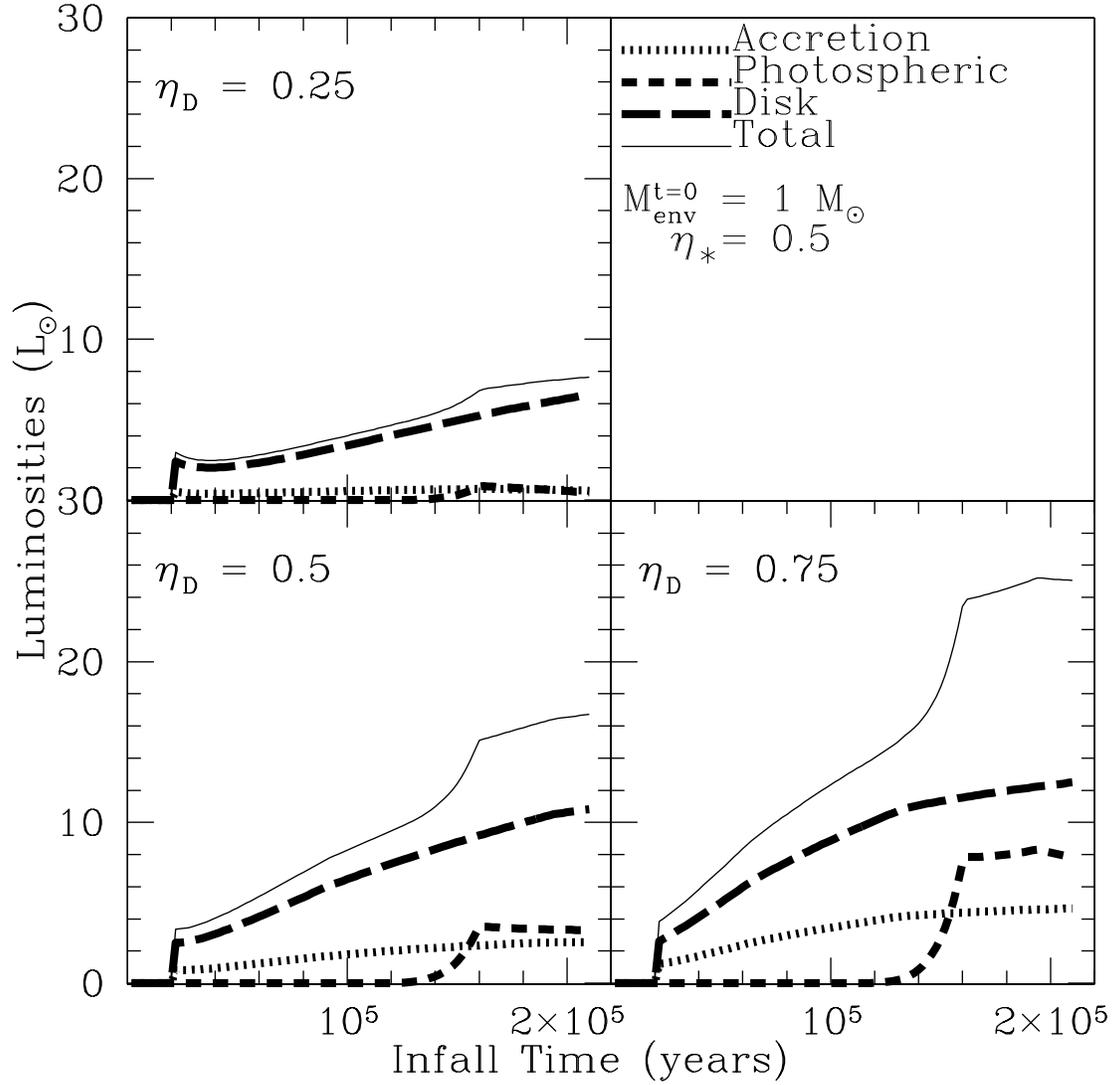} \figcaption{\label{lum_etad} Contributing forms
of luminosity are shown here for different values of $\eta_D$.  The
accretion luminosity ($L_{acc}$) is from material accreting onto the star.
Photospheric luminosity ($L_{phot}$) is calculated by DM94 and due to
contraction of the central star and deuterium burning. The disk luminosity
($L_D$) has several components, which are described in
section~\ref{sxn-diskluminosity}. The sum of these luminosities is shown by
the ``Total.''}
\end{figure}

We show the effects of changing $\eta_D$ for the 1 M$_\odot$ core in
Figure~\ref{lum_etad}. Varying $\eta_D$ alters the evolution of all
components of the luminosity.  Of course, this parameter is included
directly in the equations for the disk and stellar accretion luminosities.
Because $\eta_D$ determines the stellar mass, it also affects the
luminosity due to deuterium burning and contraction ($L_{phot}$).  With
$\eta_D=0.75$, each of these are higher than in the other two scenarios
($\eta_D=0.25$ \& 0.5).

\begin{figure}
\plotone{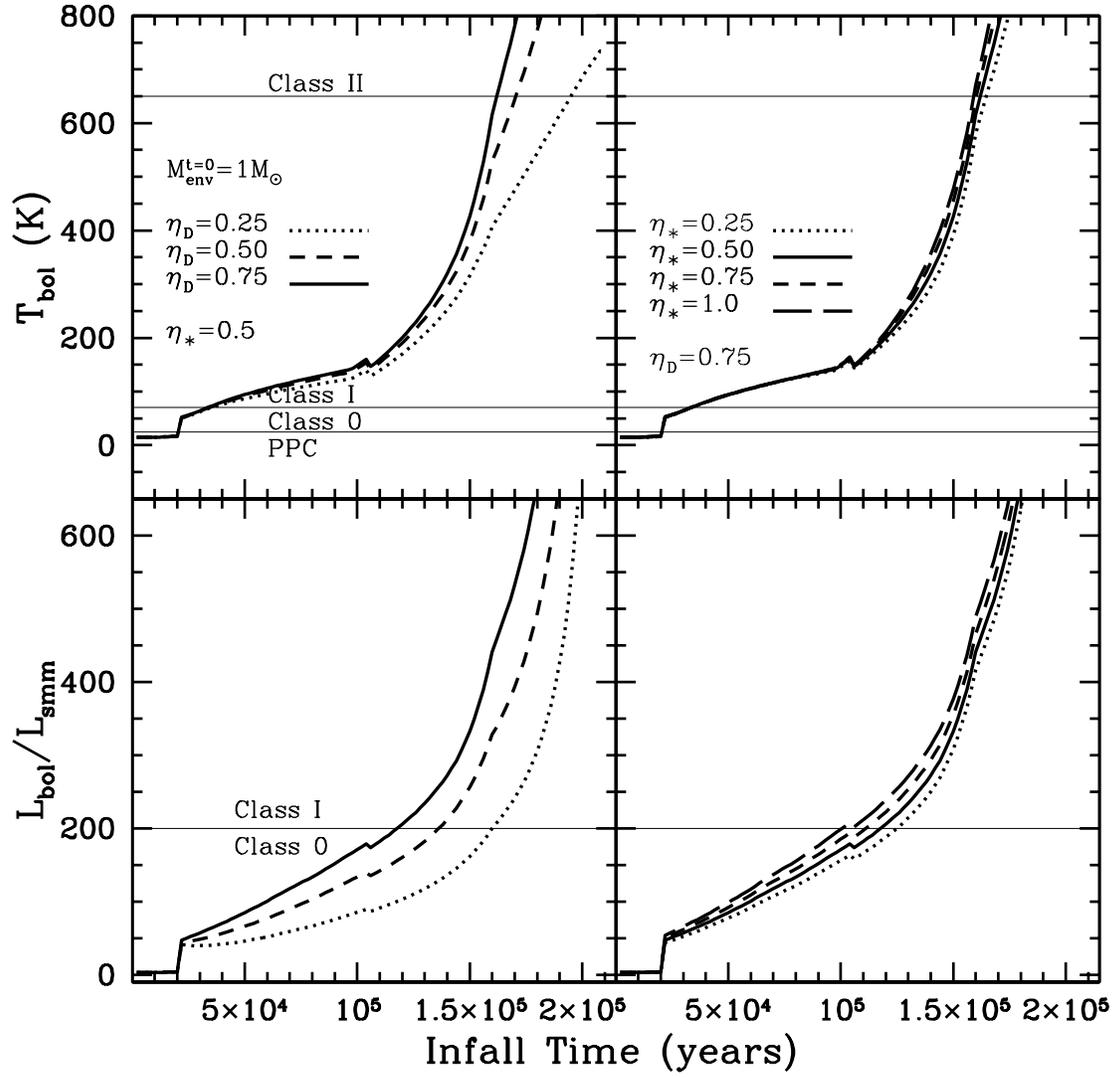}
\figcaption{\label{tbol_lsmm_eta} For the 1 M$_\odot$ core, we show
$T_{bol}$ and $L_{bol}/L_{smm}$ with different values for $\eta_D$ and
$\eta_\ast$. The left panels show data where $\eta_\ast=0.5$ and $\eta_D$
is varied; data in the right panels have $\eta_D=0.75$ and different values
for $\eta_\ast$.}
\end{figure}

Finally, in the left panels of Figure~\ref{tbol_lsmm_eta}, we show how
changing $\eta_D$ affects observational signatures.  We plot $T_{bol}$ and
$L_{bol}/L_{smm}$ versus time with $\eta_D=0.25$, 0.5, and 0.75.
Variations in $\eta_D$ have almost no effect on the evolution of $T_{bol}$,
and $L_{bol}/L_{smm}$ only varies slightly for the different values of
$\eta_D$.

\subsection{$\eta_\ast$}

\begin{figure}
\plotone{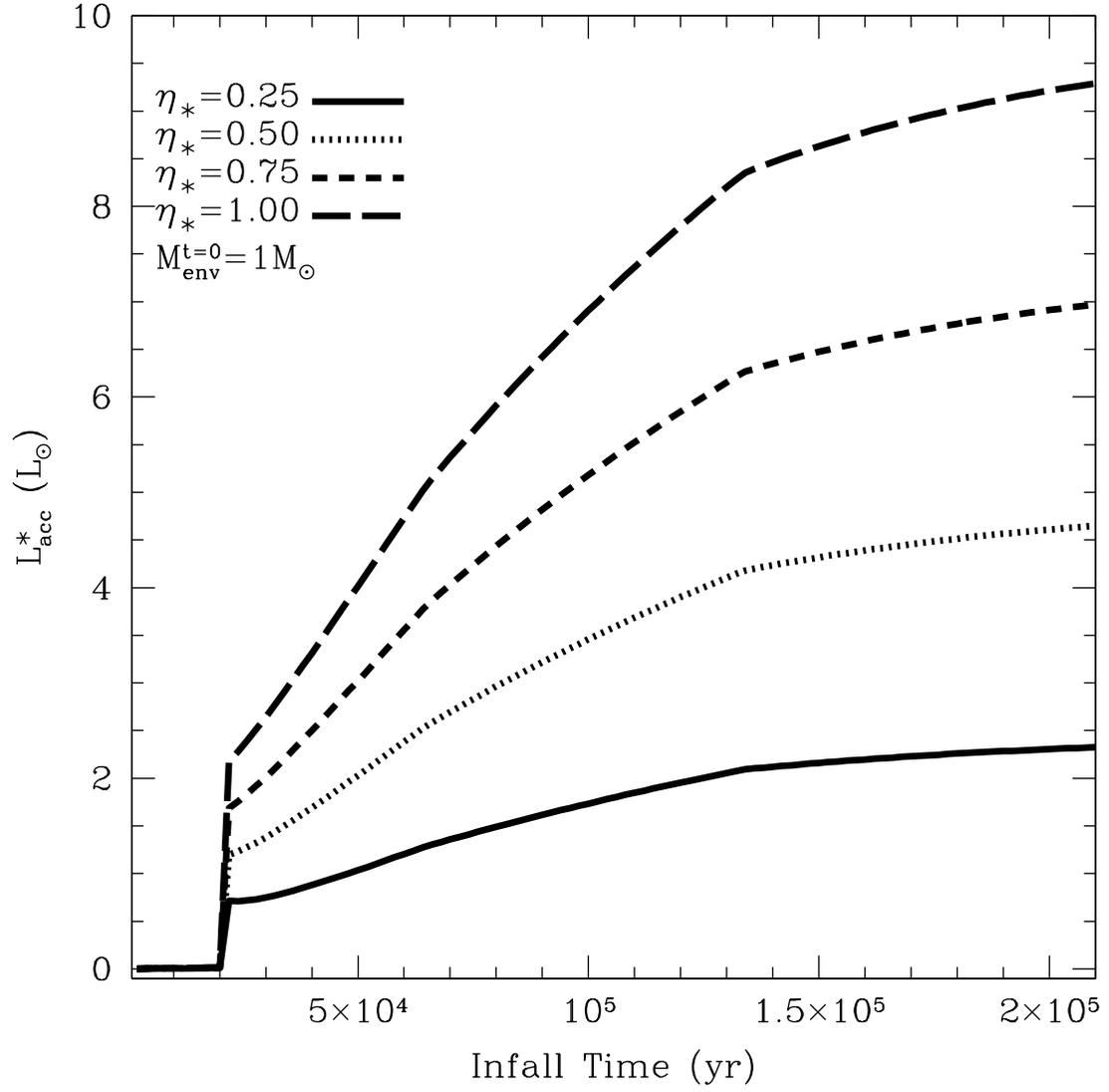} \figcaption{\label{etastar} For the 1
M$_\odot$ core, we plot $L_{acc}$ for various values of $\eta_\ast$.  For
our models, we choose $\eta_\ast=0.5$ in accord with \citet{adams87}.}
\end{figure}

This factor is a measure of how much of the total luminosity arises from
the central star.  In practice, it is only relevant for the accretion
luminosity, $L_{acc}^\ast$, as in equation~\ref{eqn-lacc-approx}.  In
Figure~\ref{etastar}, we plot the evolution of $L_{acc}^\ast$ for different
values of $\eta_\ast$.  The value that we choose for this ``efficiency
factor'' does considerably change the accretion luminosity.  However, as
shown by Figure~\ref{tbol_lsmm_eta}, the value for $\eta_\ast$ has very
little effect on the observational signatures.  For the standard model in
this paper, we adopt $\eta_\ast=0.5$.

\subsection{$\tau_{max}$}\label{sxn_tau_max}

As discussed in section~\ref{sxn-envelope}, the inner radius of the
envelope is set so that $\tau(100\mu$m) does not exceed a certain value.
Previously, some have proposed that the condition of isothermality is violated
if $\tau(100\mu$m)$>1$ \citep{larson69}.  However, \citet{masunaga99}  
showed that $\tau(100\mu$m) could be significantly greater than 1 while
the core still maintains isothermality.  

\begin{figure}
\plotone{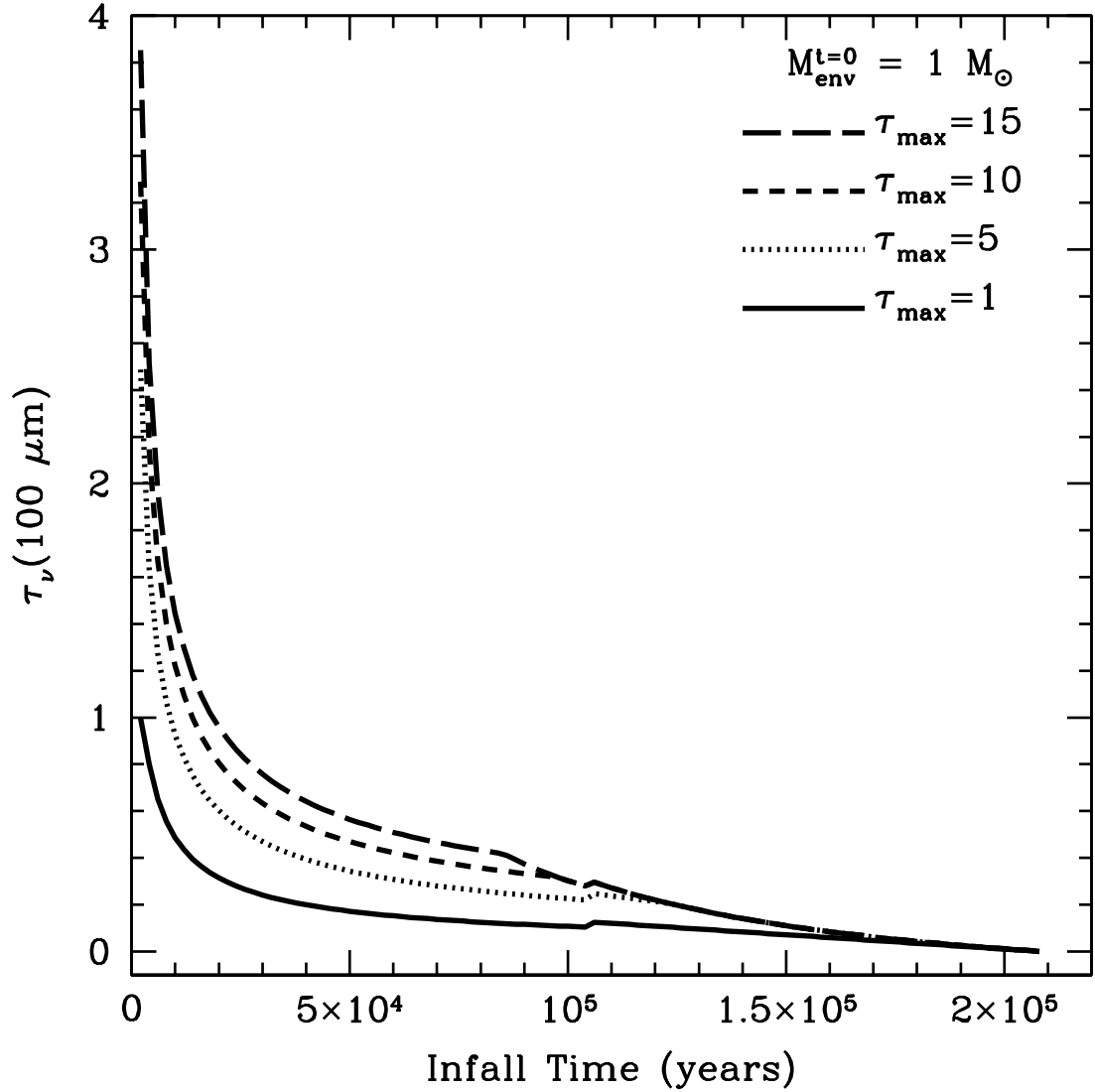} \figcaption{\label{tau} We set different
initial conditions for the forming star by allowing $\tau_\nu(100\mu$m) to
range from 1 to 15. We vary $\tau_\nu$ by changing the envelope's inner
radius.  There is a slight ``kink'' in the evolution of $\tau_\nu$ around
$t=10^5$ years that is due to the shift from the Shu77 collapsing envelope to
a free-falling envelope with $n(r)\propto r^{-3/2}$.}
\end{figure}

\begin{figure}
\plotone{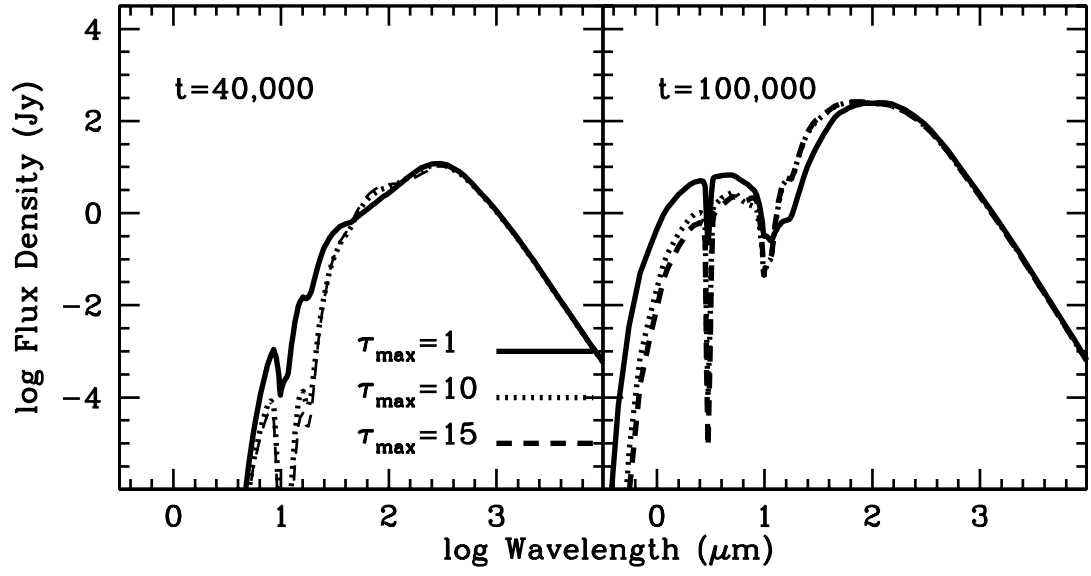} \figcaption{\label{comp} For the 1
M$_\odot$ core, we show the modeled SED with different values for
$\tau_{max}$.  The model with $\tau_{max}=1$ shows greater emission at
short wavelengths than those with higher $\tau_{max}$. We choose
$\tau_{max}=10$ for the models hereafter.}
\end{figure}

In Figure~\ref{tau}, we plot the evolution of $\tau_\nu$(100 $\mu$m) with
$\tau_{max}=1$, 5, 10, and 15.  The density profile begins with a
distribution of $n(r)\propto r^{-2}$, but, as collapse begins, the inner
region drops to a considerably shallower profile. Because the inner density
profile changes, $\tau$ drops considerably. In Figure~\ref{comp}, we show
the SED of a 1 M$_\odot$ core at $t=4\times10^4$ and $10^5$ years with
$\tau_{max}=1$, 10, and 15. The model with $\tau_{max}=1$ shows more
emission at shorter wavelengths than those with higher $\tau_{max}$.  The
SED with $\tau_{max}=10$ is almost identical, at all times, to that with
$\tau_{max}=15$.

In Figure~\ref{tau}, there is a discontinuity at $\sim1\times10^5$ years.
This is the point where the infall radius exceeds the outer radius of the
infalling envelope.  At this point, we changed from the \citet{shu77}
solution to an envelope whose density is described by $n(r)\propto
r^{-3/2}$.  This is not a perfect transition, however, so the ``kink'' at
$\sim1\times10^5$ years is an artifact of the model.

\begin{figure}
\plotone{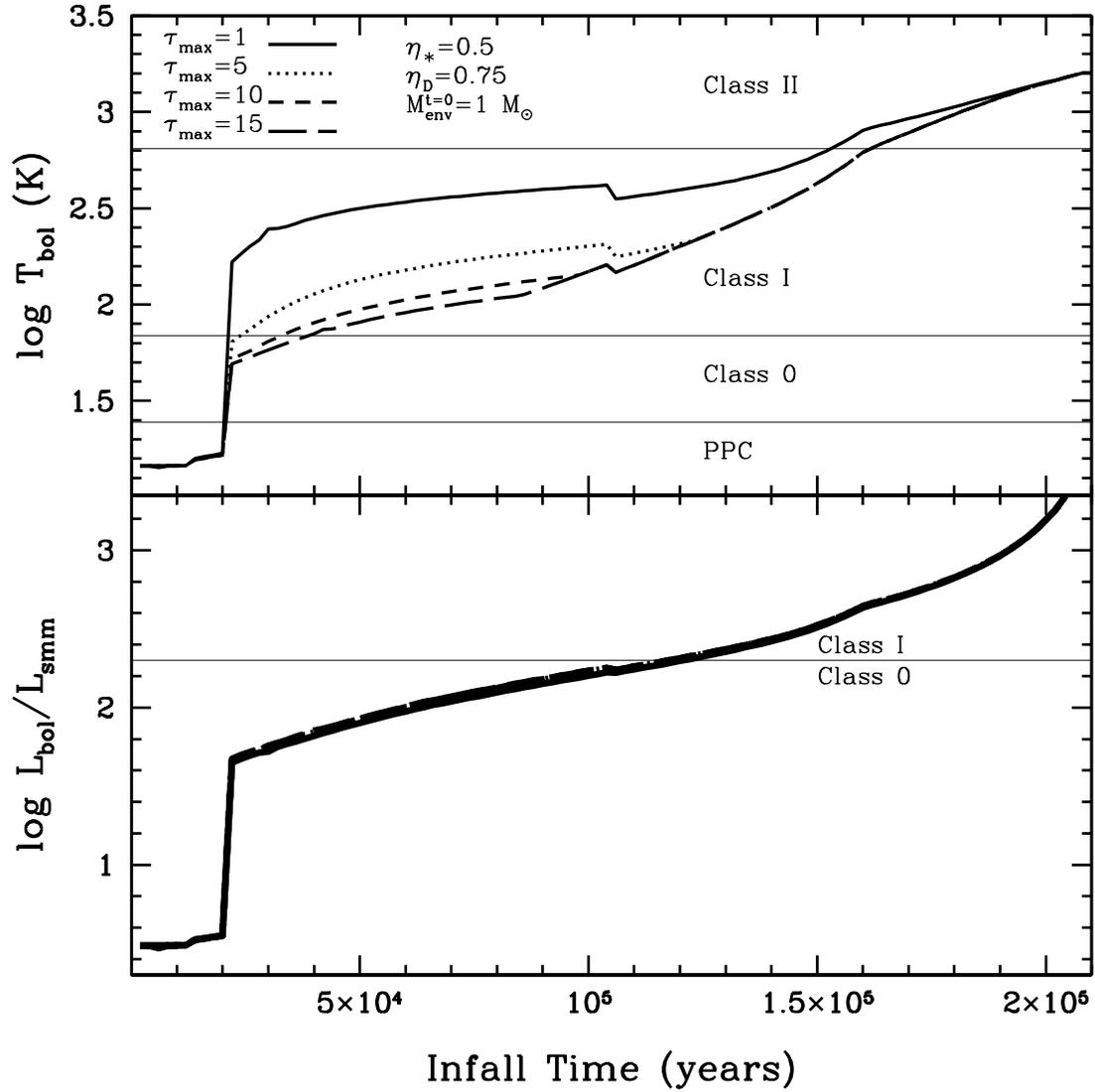}
\figcaption{\label{tbol_lsmm_tau} For the 1 M$_\odot$ core, we show
$T_{bol}$ and $L_{bol}/L_{smm}$ with different values for $\tau_{max}$.}
\end{figure}

We explore the effects of changing $\tau_{max}$ on the observational
signatures of our evolving protostar.  In Figure~\ref{tbol_lsmm_tau}, we
show the evolution of $T_{bol}$ and $L_{bol}/L_{smm}$ for the different
values of $\tau_{max}$.  The bolometric temperature is most affected by
varying $\tau_{max}$; higher values for $\tau_{max}$ slow the transition
from a Class I to a Class II protostar (as defined by $T_{bol}$).  Simply,
less short-wavelength radiation can escape the cloud when the opacity is
higher.  However, for those models with $\tau_{max}\geq 5$, the evolution
of $T_{bol}$ begins to converge.  Finally, not surprisingly,
$L_{bol}/L_{smm}$ is unaffected by changing $\tau_{max}$. We do not
significantly alter the mass of the envelope, which dictates $L_{smm}$, nor
the accretion processes, which are responsible for $L_{bol}$.

The actual value for $\tau_{max}$ is highly uncertain.  We do not fully
understand the density structure in this transition area between the
envelope and disk nor do we include the geometrical effects of a flattened
envelope in this region.  We assume that $\tau_{max}=10$; higher values of
$\tau_{max}$ have little effect on observed quantities (i.e., the SED and
its derived signatures).  The peak of a 10 K blackbody is 350 $\mu$m.
Assuming $\kappa\propto\lambda^{-1.5}$, $\tau_{350\mu m}=1.5$ when
$\tau_{100\mu m}=10$.

\section{Results}

With these free parameters set to the given values, we have run models with
different initial masses: 0.3, 1.0, and 3.0 M$_\odot$.  Then, we have 
calculated the various observational signatures. 

\subsection{$T_{bol}$ and $L_{bol}/L_{smm}$}

The bolometric temperature and the ratio of bolometric to submillimeter
luminosity have emerged, in the past decade, as the two primary methods of
classifying protostars.  Because $T_{bol}$ is a measure of the
flux-weighted mean frequency of the protostar's SED, it is highly affected
by the emergence of any short-wavelength radiation.  The ratio of the
bolometric to submillimeter luminosity, on the other hand, is virtually
unaffected by the short-wavelength emission.  Therefore, $L_{bol}/L_{smm}$
is less susceptible to the effects of geometry, which can cause more or
less NIR radiation to be observed.  This ratio is a rough measure of the
ratio of protostellar mass (including the disk and star) to the envelope
mass.

\begin{figure}
\plotone{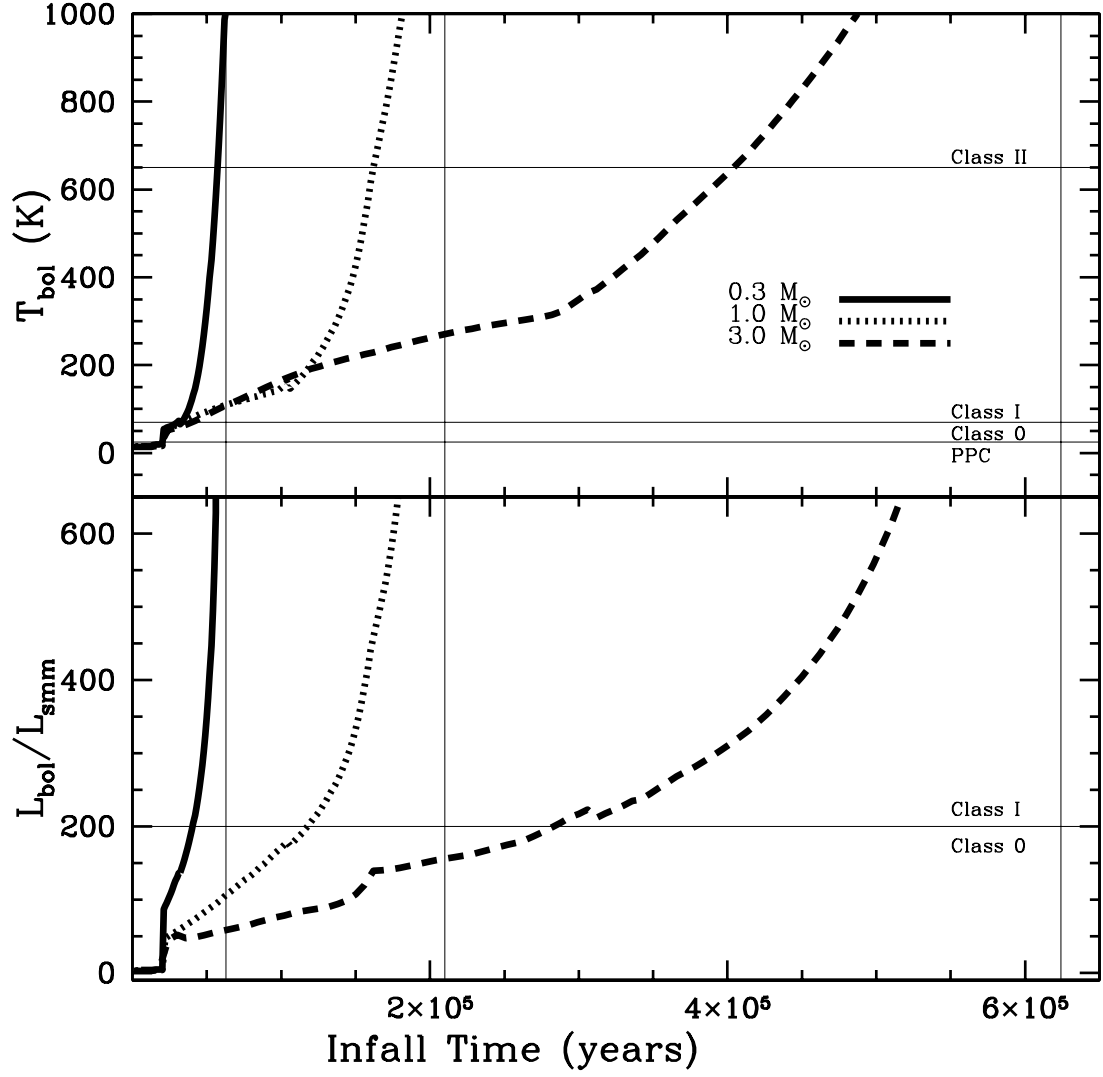}
\figcaption{\label{tbol_lsmm_mass} We plot $T_{bol}$ and $L_{bol}/L_{smm}$
as they evolve with time for each of the three models. The horizontal lines
represent the boundaries for classes of protostars; vertical lines mark
the point when all envelope material has accreted onto the star+disk system
for each of the three modeled cores.}
\end{figure}

In Figure~\ref{tbol_lsmm_mass}, we plot the two signatures as they change
for the evolving protostar. For the 0.3 M$_\odot$ core, each of the
evolutionary indicators increase drastically at about $2.0\times10^4$
years, when the FHC contracts.  All of the envelope material has accreted
onto the star and disk by $6.3\times10^4$ years, shortly after the central
star has contracted from the FHC stage.  Therefore, as the central star
becomes hotter and more luminous, there is little material left in the
envelope.  Then, $T_{bol}$ increases because more short-wavelength
radiation is observed, and $L_{bol}/L_{smm}$ increases because
$L_{smm}\rightarrow 0$ as the envelope goes away.  If this model is
correct, these low-mass stars should proceed through the FHC stage and,
almost immediately, be seen as Class II objects.

The 1.0 and 3.0 M$_\odot$ objects track one another fairly well, up to
$10^5$ years, in the $T_{bol}$ plot (Figure~\ref{tbol_lsmm_mass}) despite
the fact that they form on different timescales.  The 1.0 $M_\odot$ core
requires $2.1\times10^5$ years for all envelope mass to accrete while the
3.0 $M_\odot$ core requires $6.24\times10^5$ years. However, both cores
evolve from Class 0 to Class I at about 50,000 years, which is about $1/4$
and $1/10$ of the total infall time for the 1.0 and 3.0 M$_\odot$ cores,
respectively. 

This transition from Class 0 to I is partly due to the sudden ``turning
on'' of the central source as it contracts to $\sim3$ R$_\odot$, and
accretion luminosity becomes relevant.  There is, however, another reason
for this sudden transition.  As shown in Figure~\ref{tau}, the Shu77 model
exhibits a drastic decrease in $\tau$ regardless of the adopted value for
$\tau_{max}$.  Initially, the envelope is described by $n(r)\propto
r^{-2}$, but, as collapse ensues, it changes to $r^{-3/2}$ and $\tau$
drops.  Because $\tau$ is so low, any substantial source of stellar
luminosity will cause observable short-wavelength radiation to emerge from
the system.

These details are actually quite important if one uses these evolutionary
signatures to derive relative lifetimes of the various classes as has been
done in the past.  If one calculates the bolometric temperature for a group
of protostars (whose SEDs have been well sampled), there should be very few
Class 0 cores---conservatively, about 1/10 to 1/4 of the Class I
population, but most likely a much smaller fraction.  \citet{visser02}
presented their efforts to do such a study; they found approximately equal
numbers of Class 0 and I objects.  However, their data only included the
far-infrared (IRAS) and submillimeter fluxes.  Analysis of the mid- and
near-infrared data from the Spitzer Legacy and 2MASS surveys along with
far-infrared and millimeter observations will almost certainly produce
different results.  With more complete sampling of the protostars' SEDs,
very few Class 0 cores, by the $T_{bol}$ criterion, should remain if this
picture of evolution is correct.

Finally, in the plot of $T_{bol}$, the 0.3, 1.0, and 3.0 M$_\odot$ data
exhibit a ``kink'' at about $3\times10^4$, $10^5$, and $3\times10^5$ years,
respectively.  This is the point where the envelope's density distribution
is described by a power-law, $n(r)\propto r^{-3/2}$, instead of the Shu77
solution.  The power-law distribution has a slightly higher $\tau$ that
causes less short-wavelength radiation to be observed.  As a result,
$T_{bol}$ decreases slightly at this point of transition, but this is an
artifact of the model.

The ratio of bolometric to submillimeter luminosity seems to be much more
consistent in describing the evolution of these protostars.  Adopting
$L_{bol}/L_{smm}=200$ as the dividing line for Class 0 and I cores, we find
that the 1 and 3 M$_\odot$ protostars become Class I objects after
$1.18\times10^5$ and $3.6\times10^5$ years, respectively.  These times
correspond to slightly more than 1/2 of the total infall time for each core
whereas the $T_{bol}$ criterion showed that the cores became Class I after
1/4 and 1/10 of their total infall times.

\begin{figure}
\plotone{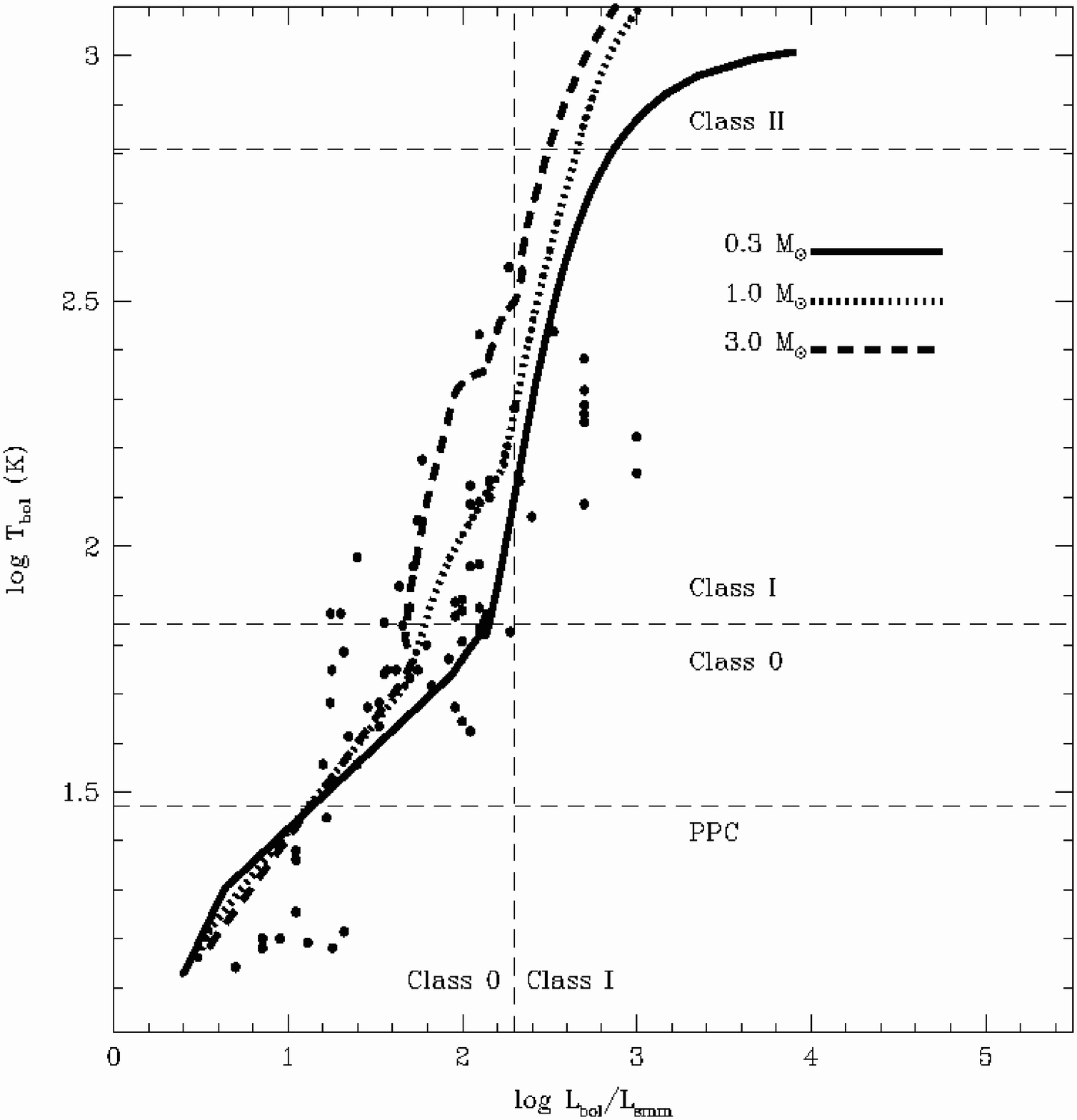} \figcaption{\label{tbol_lsmm} We plot
$T_{bol}$ and $L_{bol}/L_{smm}$ for the three scenarios that we modeled.
The points are data taken from \citet{young03}, \citet{shirley02},
\citet{shirley04}, and \citet{froebrich05}.}
\end{figure}

Finally, in Figure~\ref{tbol_lsmm}, we show a plot of $T_{bol}$ and
$L_{bol}/L_{smm}$ for the three models.  The points represent data from
\citet{young03}, \citet{shirley00}, and \citet{froebrich05}.  In general,
these models are consistent with the data.  However, the 11 starless cores
in the lower left-hand section of this plot show higher $L_{bol}/L_{smm}$
than the model predicts.  The definition of $L_{bol}/L_{smm}$ includes data
longward of 350 $\mu$m, but, for all of these cores, no 350 $\mu$m data
exist.  Therefore, the observed $L_{smm}$ is lower than that which is
modeled.  Second, there is little near- or mid-infrared data available for
many of the sources represented.  With future observations, the bolometric
temperature will almost certainly increase.  For example, we consider the 1
M$_\odot$ model and calculate $T_{bol}$ by including different fluxes.  If
we use IRAS fluxes only, $T_{bol}=83$ K for the core at 5$\times10^4$ years
while $T_{bol}$, calculated with just the Spitzer bands, is 92 K. At
$t=10^5$ years, $T_{bol}$ is 88 and 151 K as measured with the IRAS and
Spitzer fluxes, respectively.

\subsection{Mass Ratio}\label{sxn-massratio}

\begin{figure}
\plotone{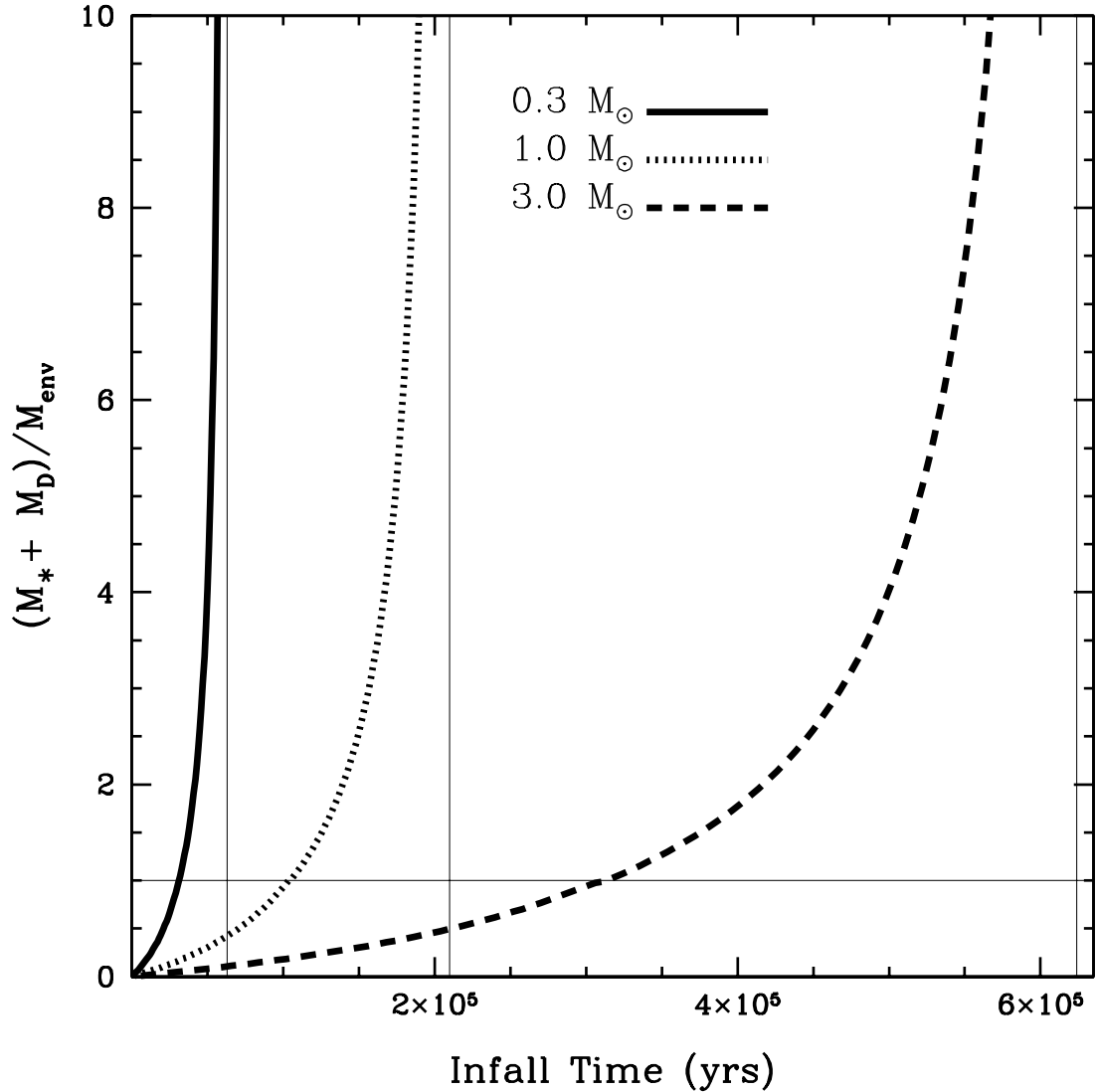} \figcaption{\label{mratio_env} For
the three models, we plot $(M_\ast+M_D)/M_{env}$.  The thin, vertical lines
mark the point when all of the envelope mass has accreted onto the
star+disk system. The horizontal line shows where the envelope mass is
equal to the accreted protostellar mass.}
\end{figure}

We can look at classification from a different, more physical, perspective.
In Figure~\ref{mratio_env}, we plot the ratio of the stellar and disk mass
to the envelope mass.  Physically, we might consider a protostar to move
from Class 0 to Class I when this ratio is 1 and there are equal amounts of
mass in the protostellar system and the envelope surrounding this
star+disk.  This event occurs at $t=3.1\times10^4$, 1.05$\times10^5$, and
3.1$\times10^5$ years for the 0.3, 1.0, and 3.0 M$_\odot$ cores.  Of
course, as defined by Shu77, this is also the time when the infall radius
is equal to the outer radius.

\begin{figure}
\plotone{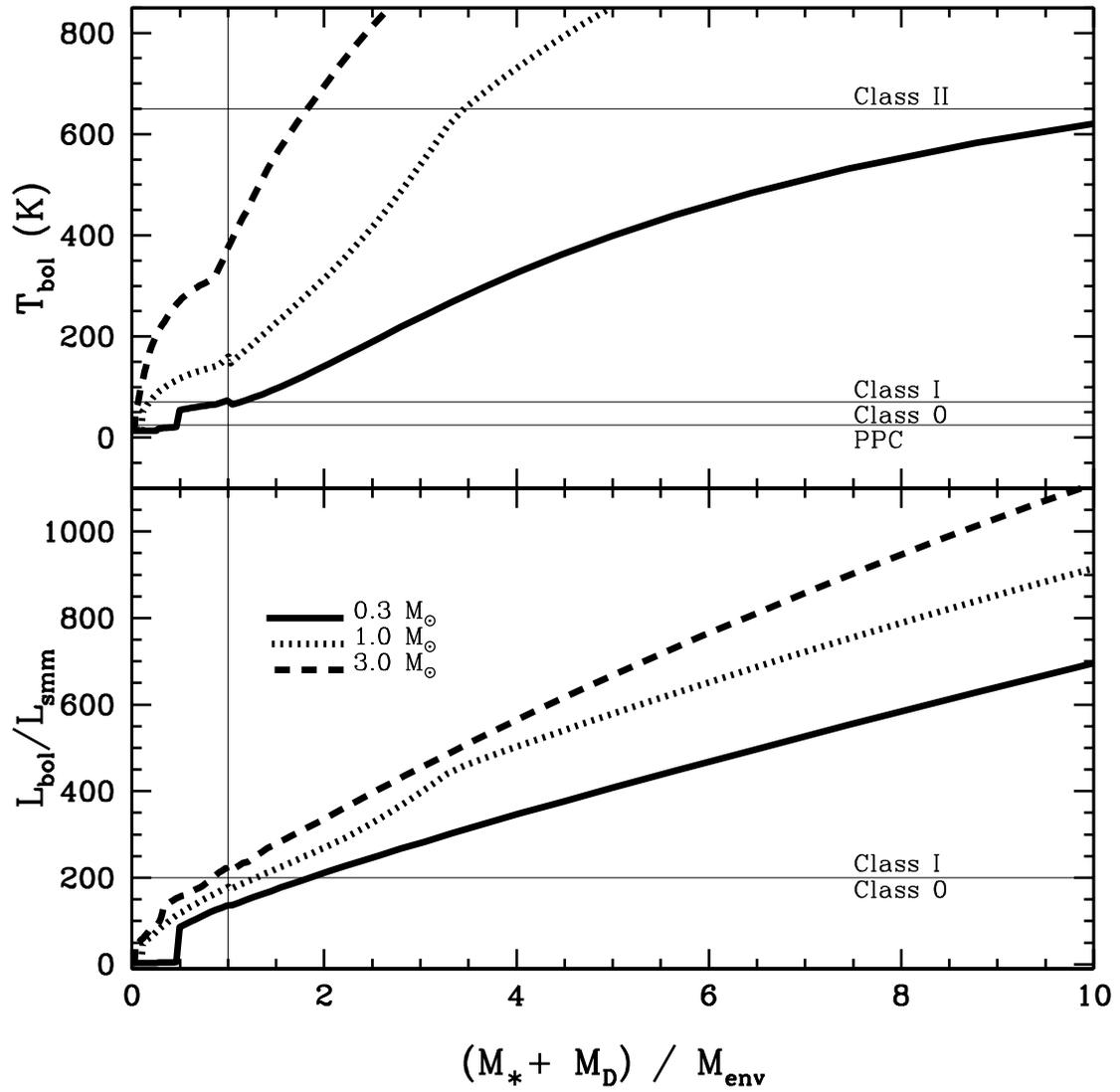}
\figcaption{\label{tbol_lsmm_mratio} Two evolutionary signatures, $T_{bol}$
and $L_{bol}/L_{smm}$, are plotted against the ratio
$(M_\ast+M_D)/M_{env}$.  The thin horizontal lines denote boundaries for
the class transitions while vertical line marks where the star+disk mass
equals the envelope mass.}
\end{figure}

In Figure~\ref{tbol_lsmm_mratio}, we plot $T_{bol}$ and $L_{bol}/L_{smm}$
as a function of $(M_\ast+M_D)/M_{env}$.  In the $T_{bol}$ plot, the 1 and
3 M$_\odot$ cores change from Class 0 to Class I while
$(M_\ast+M_D)/M_{env} < 0.5$.  With the presently defined boundaries, the
bolometric temperature does not appropriately classify the stages of star
formation. Only the 0.3 M$_\odot$ core changes from Class 0 to I when
$(M_\ast+M_D)/M_{env}\sim1$ as is appropriate for our understanding of
these stages.

However, with the $L_{bol}/L_{smm}$ criterion, we find that the cross-over
from Class 0 to I occurs approximately when $(M_\ast+M_D)/M_{env} = 1$,
which is a more realistic view of these evolutionary stages. Indeed, this
observational signature is also favored because it is not largely
dependent on what is observed at short wavelengths where geometric effects
play a big role \citep{andre93}.

Therefore, we set some physical divisions for the evolutionary transitions.
First, we let the PPC/Class 0 transition occur when the FHC first collapses
at $\sim2.0\times10^4$ years.  For all three cores, this occurs
when $L_{bol}/L_{smm} \sim 35$.  For the Class 0/I transition, we let
$(M_\ast + M_D)/M_{env}=1$; $L_{bol}/L_{smm} \sim 175$ when this criterion
is met.  Notice that this value is slightly less than the requirement given
by \citet{andre93}, i.e.  $L_{bol}/L_{smm}=200$. The Class II stage begins
when all of the envelope material has been accreted.  Our models are not
reliable at these late times, and the $L_{bol}/L_{smm}$ signature is not a
very good indicator for this stage of evolution.  Nonetheless, we find that
$L_{bol}/L_{smm}$ is approximately 2000 at this point, but this value is
dependent on the adopted model for the disk.

\subsection{BLT Diagrams: A Comparison with M98}

\begin{figure}
\plotone{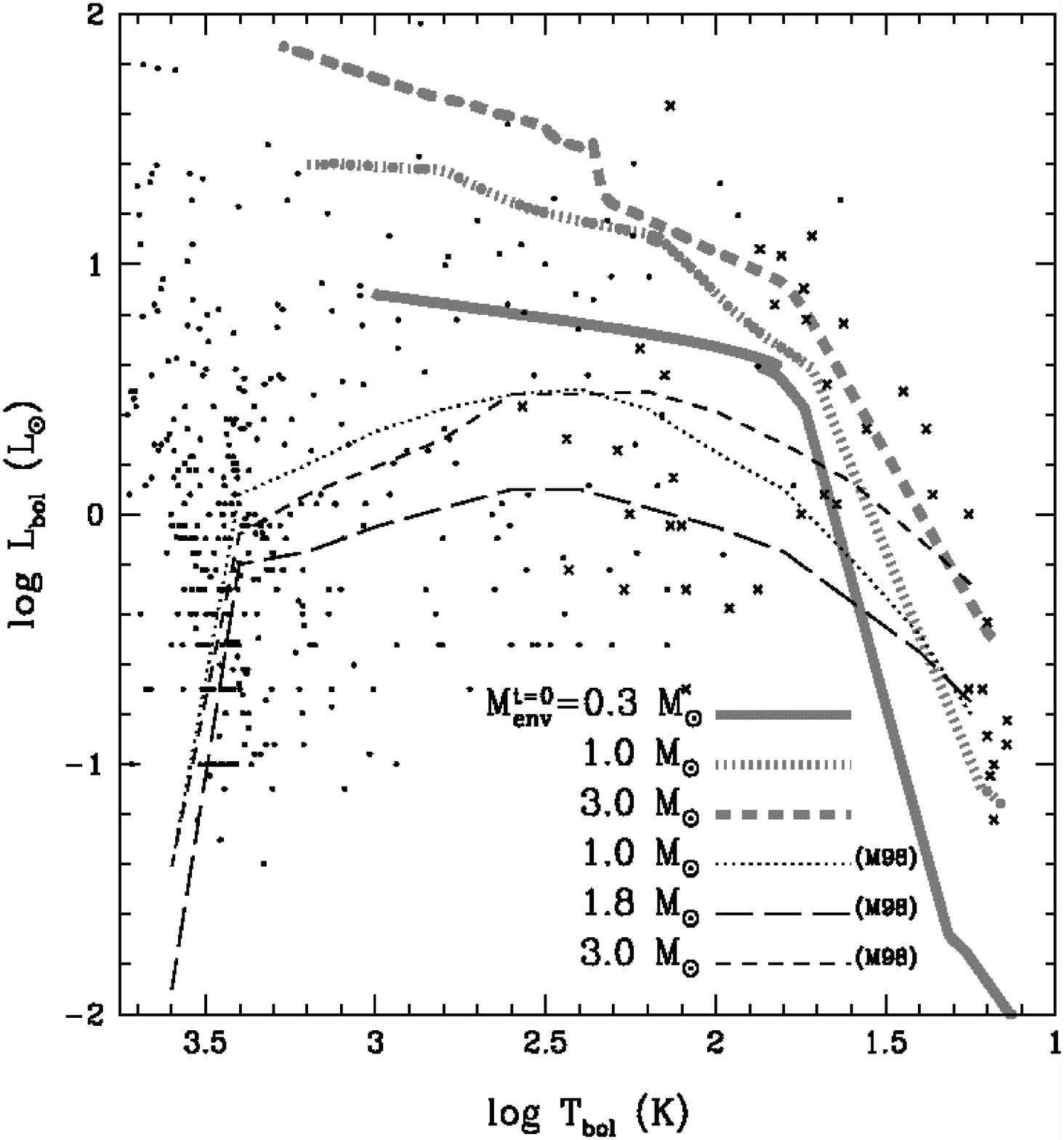} \figcaption{\label{blt} As seen in M98, this is a
bolometric luminosity and temperature diagram.  The thick lines are our
models for three different initial masses.  The thin lines are models from
M98 with the initial envelope masses labeled.  The 1.0 and 3.0 M$_\odot$
models are from figure 7 in M98. The 1.8 M$_\odot$ model is from figure 9
in M98.  The crosses are from \citet{young03}, \citet{shirley02}, and
\citet{shirley04}. The dots are from \citet{chen95} and \citet{chen97}.}
\end{figure}

In Figure~\ref{blt}, we show a plot of the bolometric luminosity and
temperature, which M98 called a BLT diagram.  The axes are laid out to
mimic the Hertzsprung-Russell diagram with $T_{bol}$ increasing right to
left.  In Figure~\ref{blt}, we have included M98's models from Figure 7 in
their paper.  Two of the thin lines in Figure~\ref{blt} are their models
for forming, 0.5 M$_\odot$ protostars whose initial envelope masses were 1
and 3 M$_\odot$.  We also show their 0.3 M$_\odot$ model from Figure 9, but
we label it here as 1.8 M$_\odot$ because this is the mass of the envelope
before collapse begins while 0.3 M$_\odot$ is the mass of the star at
t=$\infty$.  We also plot data from \citet{young03}, \citet{shirley02}, and
\citet{shirley04} as crosses; the dots represent data from \citet{chen95}
and \citet{chen97}.  Our tracks are markedly different from any of those
presented in M98, so a summary of the differences between two models is
relevant here.  Primarily, our methods differ in that M98 attempts to
create a reasonable model that fits the data while we are simply
determining the observational signatures of the Shu77 model.

The most significant difference between this work and that of M98 is the
assumptions for infall evolution.  M98 described the infall and accretion
with an exponential decay function such that the accretion rates began at
about $10^{-6}$ M$_\odot$ yr$^{-1}$ and, as the star approached the main
sequence, finished with $10^{-9}$ M$_\odot$ yr$^{-1}$.  We assume constant
accretion onto the star+disk system throughout the duration of the life of
the envelope as predicted by the Shu77 collapse solution. However, in our
model, the rate of accretion onto the star's surface does decrease with
time as the disk forms and takes a more prominent role in processing
material from the envelope to the star.  Also, the modeled evolution is
longer for M98 ($10^6$ years) than in our model ($2-6\times10^5$ years).

These different assumptions about infall have several implications.  First,
in our model, we form a more massive star from similar initial conditions
in less time.  For our 3 M$_\odot$ core, the star reaches 0.5 M$_\odot$ at
$t=132,000$ years.  On the other hand, M98's models require, by design,
about $10^6$ years to create this 0.5 M$_\odot$ star from a 3 M$_\odot$
core.  Of course, at the end of $10^6$ years, the star created by M98 has
completed its pre-main-sequence evolution, while our model still requires
time to completely accrete the disk material.

This discrepancy in the star's final mass presents another difference
between this work and that of M98.  M98 included a dispersal timescale for
the envelope that included an assumption of mass loss due to outflow from
the central protostar.  We do not include outflows in any way in these
models.  The lack of outflows in this work has two implications: 1) the
mass evolution, as depicted in Figure~\ref{mass}, is incomplete and 2) we
do not consider the effects of an evacuated outflow cavity.  Fortunately,
the mass evolution is not considerably altered by the exclusion of
outflows. \citet{calvet98} found that the ratio of mass loss rate to mass
accretion rate is $\sim 0.1$.  We are unable to model the scattered light
coming from the outflow cavity.  Others have, however, and we discuss their
work in Section~\ref{sxn-whitney}.

M98 also assume the envelope to have a density profile with a free-falling
structure, $n(r)\propto r^{-3/2}$.  We use the Shu77 solution, which has an
inner free-falling envelope surrounded by a static envelope with
$n(r)\propto r^{-2}$.  These differences in structure of the envelope cause
great disparities in the opacity.  For example, a core that has 0.8 M$_\odot$
of material with $n(r)\propto r^{-3/2}$ creates $\tau_\nu(100\mu m)=0.26$.
A core with the same amount of material but described by the Shu77 collapse
solution, with an infall radius that is one-half of the outer radius,
creates $\tau_\nu(100\mu m)=0.17$.  Such a disparity causes large changes
in $T_{bol}$.  For example, if we place a 0.7 L$_\odot$ star with
T$_{eff}$=2000 K inside these two cores, the Shu77 core has $T_{bol}=416$ K,
and the free-falling core has a bolometric temperature that is half as high,
$T_{bol}=207$ K.

M98 do not calculate the full SED.  Instead, they consider the optically
thin and thick limits and calculate two moments of the protostellar
spectrum: the bolometric temperature and luminosity.  On the other hand, we
use DUSTY to calculate the full radiative transfer in the protostellar
system.

Finally, M98 use a single power-law to describe the dust emissivity,
$\kappa$.  In contrast, we use the dust properties calculated by
\citet{ossenkopf94}. While the dust opacity is aptly described by a
power-law at long wavelengths, the shorter wavelength opacities are clearly
not properly represented in the same way (see Figure~\ref{kappa}).
  
Interestingly, the data encompass both models, but M98's models obviously
best cover the median range for the data.  However, there are a substantial
number of sources that have a lower bolometric luminosity than either model
allows.  Perhaps, these sources are in the quiescent stage of episodic
accretion so that $L_{acc}$ is very low, or these low-luminosity objects
could simply have a mass less than 0.3 M$_\odot$ and not be included in
these models. Also, our models show higher luminosities at later times than
most of the data, a result of the aforementioned ``luminosity problem.''
Perhaps, accretion does occur episodically and only for short times, and we
should expect only a few objects to be in the phase where material is
accreting onto the star and, hence, have a high luminosity.  Another
explanation, of course, is that the assumption of constant accretion
throughout the star's evolution is wrong.

\subsection{Infrared Color-Magnitude Diagrams}\label{sxn-whitney}

\begin{figure}
\plotone{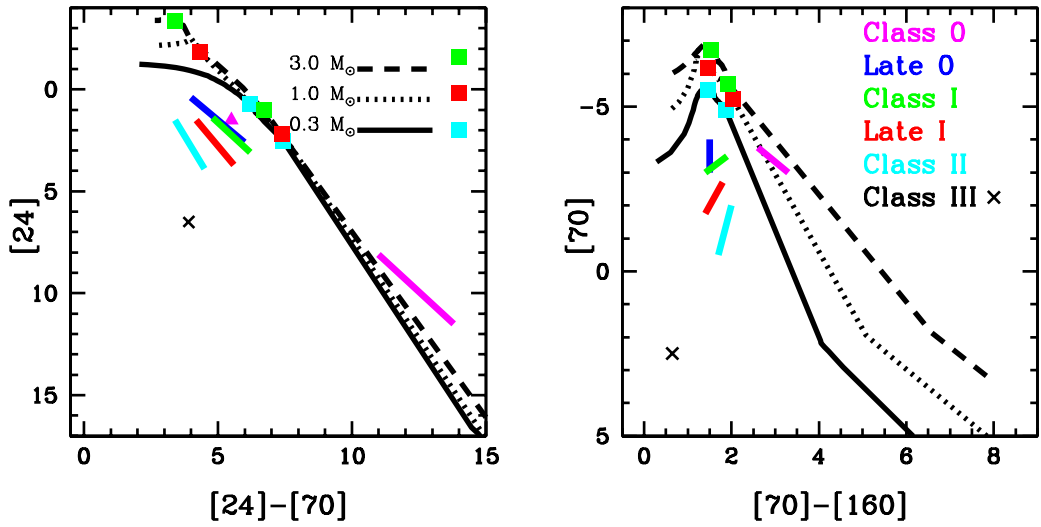}
\figcaption{\label{sirtf_colors} The black lines in this plot represent the
SST MIPS colors as calculated for our models; $[24]$, $[70]$, and $[160]$
are the magnitudes observed with MIPS.  The colored lines are from Whitney
et al. (2003) and extend over varying inclination angles.  For the Class 0
models, the line represents most inclination angles except for the pole-on
scenario, which is shown here as a magenta, filled triangle.  Transitions,
as discussed in Section~\ref{sxn-whitney}, for the 1 M$_\odot$ models are
shown as red squares while the green squares represent the transitions for
the 3 M$_\odot$ model and cyan for the 0.3 M$_\odot$ core.}
\end{figure}

In Figure~\ref{sirtf_colors}, we show color-magnitude diagrams as would be
observed with the Multiband Imaging Photometer for Spitzer (MIPS) on the
Spitzer Space Telescope (SST).  The three mass sequences (0.3, 1.0, and 3.0
M$_\odot$) are represented by the black lines in these figures.  We show
the magnitude at 24 $\mu$m ($[24]$) plotted against the $[24]-[70]$ color;
on the right of Figure~\ref{sirtf_colors}, we plot $[70]$ and $[70]-[160]$.

Also, we show the models of W03 from Figure 8a of their work.  The colored
lines show W03's calculations over varying angles of inclination.  The
magenta line represents almost all inclination angles for the Class 0
stage, and the magenta triangle is the Class 0 stage as viewed pole-on.

Many differences exist between these models and our work.  For example, W03
describe the envelope as a rotating, freely falling envelope
\citep{ulrich76}; W03 also set the envelope's inner radius to be quite
small ($\sim10$R$_\odot$) while we use a much larger inner radius.  In
addition, W03 have a number of ``common'' model parameters, which do not
change for the different models.  These parameters include the stellar
radius, temperature, and mass as well as the overall source luminosity.  In
our model, we allow all of these to evolve with time.  Finally, W03
included the effects of outflow cavities, 2-dimensional disks, and varying
inclination angles.  We are unable to account for these things.

Young objects in the PPC/Class 0 segment of their lifespan are very red in
our calculations and occupy the lower, right-hand section of this plot.
However, these models are bluer in this plane than the W03 models for a
Class 0 source.  Several factors probably contribute to this disparity.
Most of the difference arises from the fact that W03's models have a
considerably more luminous central source than our models.  Thus, the
envelope material is heated more and emits a greater 70 $\mu$m flux.
Because W03 uses a very small inner radius, the optical depth is larger
than for our models even though they may have similarly massive envelopes.
Therefore, relative to the 24 $\mu$m flux, there is a higher 70 $\mu$m flux
for W03's models, and their Class 0 stage appears redder (and, of course,
more luminous).  The effect is much less pronounced in the $[70]-[160]$
plane because these longer wavelength fluxes are less affected by optical
depth effects.

As with W03's models, our evolutionary tracks show that the objects become
bluer in the $[24]-[70]$ color as the source becomes more luminous and
$[24]$ increases.  We have marked the PPC/Class 0 and Class 0/I transitions
(as determined by $L_{bol}/L_{smm}$ as in section~\ref{sxn-massratio}).
The Class I/II transition occurs at the end of our modeling sequence when
all envelope mass has accreted onto the star and disk.  However, aspherical
effects are most relevant with these later stages, so these models probably
underestimate the 24 $\mu$m flux for the Class I/II transition.
Transitions for the 0.3 M$_\odot$ models are shown as cyan squares, the red
squares represent the transitions for the 1 M$_\odot$ model, and the green
squares are for the 3.0 M$_\odot$ model.

The right panel of Figure~\ref{sirtf_colors} shows a simpler behavior for
the evolution as seen at 70 and 160 $\mu$m.  As an FHC, the protostar
appears very red but becomes bluer as the core collapses to $\sim$3
R$_\odot$ and grows in luminosity.  The 160 $\mu$m flux increases over time
as the internal luminosity continues to grow.  However, there is a point
where the envelope contains so little material that the 160 $\mu$m flux
drops despite the fact that the far-infrared emission of the internal star
is growing.  By the end of these tracks, the objects have reached the end
of their Class I stage; they no longer have envelope material but do have
an optically thick disk.

\section{Conclusions}

We have presented the results from modeling the evolution of protostars
with 3 different masses: 0.3, 1.0, and 3.0 M$_\odot$.  The framework for
the evolution of these protostars was taken mostly from the work of
\citet{adams86} but also used the results from several other authors.
These efforts were similar and complementary to the work of \citet{myers98}
and \citet{whitney03} but employed different methods and theories and, hence,
got different results.  

We note that the evolution of these modeled protostars is significantly
affected by the existence and lifetime of the first hydrostatic core. The
work done heretofore is useful \citep{boss95,masunaga98}, but more detailed
theoretical work is needed.  As we begin to observe earlier stages of star
formation \citep{young04a}, the role of the FHC must be ascertained.

We find that the Class 0 stage, when determined by $T_{bol}$, should be
short-lived, lasting only about 1/10 to 1/4 of the protostar's life.  This
result is somewhat model dependent, but it should also extend to other
models of star formation.  Therefore, the surveys being conducted with the
Spitzer Space Telescope and 2MASS should reveal a small fraction of Class 0
to Class I objects.  However, we suggest not using the $T_{bol}$ criterion
for classification.

We find that the bolometric temperature is a poor discriminator for
protostars at early evolutionary stages.  Instead, we suggest using the
$L_{bol}/L_{smm}$ signature proposed by \citet{andre93}.  Based on physical
grounds, we suggest these boundaries for classifying protostars:
$L_{bol}/L_{smm}=35$ for PPC/Class 0, $L_{bol}/L_{smm}=175$ for Class 0/I,
and $L_{bol}/L_{smm}\sim 2000$ for the Class I/II transition. The latter is
largely dependent on the adopted disk model.  Also, $L_{bol}/L_{smm}$ is
not relevant after all envelope mass has been accreted (beyond the Class I
stage) since it is a measure of the stellar to envelope mass ratio
\citep{andre93}.

We have presented several observational tools by which the inside-out
collapse model can be effectively tested: infrared color-magnitude
diagrams, plots of the bolometric luminosity and temperature, and a plot of
$T_{bol}$ and $L_{bol}/L_{smm}$.  Large surveys will produce these
observable quantities for hundreds of young protostars and test the theory
of inside-out collapse.

\section{Acknowledgements}

We thank Minho Choi for the use of his program to calculate the density
profiles.  Also, many thanks to Moshe Elitzur, Maia Nenkova, and \v{Z}eljko
Ivezi\'{c} for providing the modified version of DUSTY that allows heating
by the ISRF and for much assistance in the use of DUSTY.  Many thanks also
to our anonymous referee whose thorough reading and thoughtful critique
have made this a better paper. This work is supported by NASA grants
NAG5-10488 and NNG04GG24G.

\clearpage

\begin{deluxetable}{lcl}
\footnotesize
\tablecolumns{3}
\tablecaption{Model Parameters\label{parameters}}
\tablewidth{0pt} 
\tablehead{
\colhead{}                &
\colhead{Symbol}                 &
\colhead{Adopted value}                 
}
\startdata 
Envelope parameters && \\
\tableline
Envelope inner radius		& $r_i$& Equation~\ref{eqn-rienv}\\
Envelope outer radius		& $r_o$& Equation~\ref{eqn-menv}\\	
Envelope dust opacity		& $\kappa_\nu$& \citet{ossenkopf94}\\
\tableline\tableline
Disk parameters && \\
\tableline
Inner radius			& $R_i$& Equation~\ref{eqn-diskinner}\\
Outer radius			& $R_o$& Equation~\ref{eqn-diskouter}\\
Accretion luminosity		& $L_{D}$& Equation 33b, \citet{adams86}\\
Temperature power-law		& $T_{D}$& Equation~\ref{eqn-tdisk}\\
\tableline\tableline
Stellar parameters && \\
\tableline
Accretion luminosity 		& $L_{acc}$& Equation 33a, \citet{adams86}\\
Photospheric luminosity		& $L_{phot}$& \citet{dantona94} \\
Effective temperature		& $T_{eff}$& Equation~\ref{eqn-teff} \\
Radius				& $R_\ast$& \citet{palla91} \\

\enddata
\end{deluxetable}

\end{document}